\documentclass[aps,prd,preprint,amsfonts,amssymb,amsmath]{revtex4}
\usepackage{graphicx}
\usepackage{dcolumn}
\usepackage{bm}
\usepackage{amssymb}
\newcommand{\be}{\begin{equation}}
\newcommand{\ee}{\end{equation}}
\newcommand\beq{\begin{eqnarray}}
\newcommand\bea{\begin{eqnarray}}
\newcommand\eeq{\end{eqnarray}} 
\newcommand\eea{\end{eqnarray}}

\newcommand{\vev}[1]{\langle #1 \rangle}

\newcommand{\GeV}{{\text{ GeV} }}

\newcommand\hs{\hat{S}}
\newcommand\hhu{\hat{H}_u}
\newcommand\hhd{\hat{H}_d}
\newcommand\hpsi{\hat{\Psi}}

\def\mag[#1]{\left| #1 \right|}
\def\exv[#1]{\langle #1 \rangle}
\def\sl[#1]{{\displaystyle{\not} #1}}

\newcommand\susy{{\sc Susy}}

\begin{document}

\title{Dark matter in natural supersymmetric extensions of the Standard Model}
\author{Francesc Ferrer}
\email{ferrer@physics.wustl.edu}
\author{Christopher Spitzer}
\email{cspitzer@physics.wustl.edu}
\affiliation{Physics Department and McDonnell Center for the Space Sciences, Washington University, St Louis, MO 63130, USA}

\begin{abstract}
We explore the dark matter sector in extensions of the Minimal Supersymmetric 
Standard Model (MSSM) that can provide a good fit to the PAMELA cosmic ray 
positron excess, while at the same time addressing the little hierarchy 
problem of the MSSM. 
Adding a singlet Higgs superfield, S, can account for the 
observed positron excess, as recently discussed in the literature, but 
we point out that it requires a fine-tuned choice for the parameters of the 
model. 
We find that including an additional singlet, $\Psi$, allows both a reduction
of the weak-scale fine-tuning, and an interpretation of the
cosmic ray observations in terms of dark matter annihilations 
in the galactic halo. Our setup contains a light axion, but does not require 
light CP-even scalars in the spectrum.  
\end{abstract}

\maketitle

\section{Introduction}

\subsection{Dark Matter and Cosmology}

Within the standard $\Lambda$ Cold Dark Matter cosmological model ($\Lambda$CDM), 
which reproduces the available experimental observations with remarkable 
success, about a fifth of the energy density in the universe is contributed
by dark matter (DM)~\cite{Komatsu:2008hk}. 
The DM cannot be in the form of ordinary baryonic matter as deduced from
considerations of cosmological nucleosynthesis and observations of the
anisotropies of the cosmic microwave background (CMB).
Despite constituting
most of the mass in the universe and playing a crucial role in the
growth and dynamics of structure, either the lack of appreciable interactions with 
observable particles or its mass have so far prevented the determination of 
its origin and composition.

On the other hand, elementary particle theory provides several 
candidates for the DM, such as Weakly Interacting Massive Particles (WIMPs).
What makes WIMPs attractive DM candidates is that their 
existence is motivated {\em independently} in particle physics extensions 
that address some of the shortcomings of the Standard Model.
For instance, one of the by-products in extensions 
relying on low energy supersymmetry (\susy)
to alleviate the hierarchy problem in the 
higgs sector of the SM, is the natural occurrence of a stable 
WIMP. The neutralino in the Minimal Superymmetric Standard Model (MSSM) 
is the simplest viable example.
By virtue of its weak interactions, the thermal relic density 
of a WIMP is of the right order of magnitude, and its existence can be probed
through both direct and indirect detection experiments
(see e.g.~\cite{Jungman:1995df,Bertone:2004pz} for reviews).

Although no convincing signal has yet been found, the latest PAMELA
data~\cite{Adriani:2008zr,Adriani:2010ib}, suggesting a new source of galactic positrons,
has been interpreted as the result of annihilation of DM particles in 
the galactic halo~\cite{Cirelli:2008jk,Barger:2008su,Cholis:2008hb,ArkaniHamed:2008qn,Pospelov:2008jd,Nelson:2008hj,Nomura:2008ru,Harnik:2008uu,Fox:2008kb,Hooper:2009fj,Bai:2009ka,Hooper:2009gm,Park:2009cs,Wang:2009rj} or DM decay~\cite{Yin:2008bs,Nardi:2008ix,Ibarra:2008jk,Ishiwata:2009vx,Chen:2009ew,Arvanitaki:2009yb}.

On the other hand, several astrophysical sources such as pulsars~\cite{Hooper:2008kg,Yuksel:2008rf,Profumo:2008ms}, supernova remnants~\cite{Shaviv:2009bu,Blasi:2009bd}, or secondary production in regions where cosmic rays are accelerated~\cite{Blasi:2009hv,Cowsik:2009ga,Mertsch:2009ph,Ahlers:2009ae} may also 
account for
part of, or perhaps all, of the observed fluxes. In addition, the hard injection spectrum 
required to fit the positron fraction measured by PAMELA, and the
non-observation of an equivalent anti-proton excess~\cite{Adriani:2008zq} 
exclude annihilation of thermal WIMPs, 
such as the neutralino in the MSSM, as a viable explanation.

A DM interpretation of the data demands that the particles making up the
dark galactic halo annihilate mostly to charged leptons with a cross-section
which is $\sim 10-100$ times larger than the canonical value leading to
the correct cosmological abundance via thermal decoupling~\cite{Lee:1977ua}, 
$\vev{\sigma v}_{\textrm{ann}} \approx 3\times 10^{-26} \textrm{ cm}^3 
\textrm{ s}$.
MSSM neutralinos could annihilate primarily to leptons, either due to
radiative corrections in a small region of parameter 
space~\cite{Bergstrom:2008gr}, or for a wino-like LSP of about 200 GeV 
as discussed in~\cite{Grajek:2008pg}, although a non-standard cosmological
evolution has to be invoked to explain the {\em boosted} annihilation 
cross-section required by observations. A nearby clump of DM could raise
the annihilation rate~\cite{Hooper:2008kv}, but it is highly unlikely that
a sufficiently large clump can be found in the solar neighborhood
\cite{Brun:2009aj}.
The generic expectation in the MSSM is that neutralinos annihilate to a mixture of
heavy quarks and Higgs bosons producing a spectrum that is too soft to account
for the PAMELA data, while also over-producing anti-protons~\cite{Gogoladze:2009kv}.

Several alternative particle physics scenarios have been proposed that predict
DM particles can explain the rise in the positron flux at high 
energies without conflicting with other measurements. 
For instance, if the DM particles annihilate with a weak-scale strength to 
light metastable mediators 
which are very weakly coupled to the Standard Model, then
kinematical constraints preclude any final states other than light leptons.
In addition, the exchange of light mediators
results in long-range interactions that enhance the annihilation
cross-section when the DM particles are moving at non-relativistic
velocities, as is the case for present day processes occurring in the
galactic halo. This so-called Sommerfeld enhancement~\cite{Hisano:2004ds,Cirelli:2007xd,ArkaniHamed:2008qn}, 
reconciles the required underlying weak-scale interaction 
at the time of DM freeze-out that generates the correct relic abundance 
with a much larger
annihilation cross-section in the present galactic environment.
The scenarios in~\cite{Pospelov:2007mp,ArkaniHamed:2008qn,Nomura:2008ru}, among others,
provide particular implementations of these principles, but they are not
primarily motivated by a solution of the hierarchy problem.

\subsection{Natural Models of Dark Matter}

In this paper, we seek a model of dark matter that addresses the hierarchy problem and other naturalness constraints while still generating the fluxes observed by PAMELA. We require the following
\begin{itemize}
\item Stable dark matter with a mass of 100 GeV or above.
\item Dark matter annihilations dominantly to light leptons.
\item Solution to the hierarchy problem.
\item Minimal fine-tuning among model parameters.
\item Thermally generated dark matter.
\item Passes constraints from accelerators.
\end{itemize}
We find that this set of requirements will place severe restrictions on the form of the model.

We begin with a consideration of naturalness in supersymmetric models.
The MSSM, while solving some problems of the Standard Model, introduces new theoretical problems that suggest it might not be a complete description of physics at the electroweak scale.
It must contain a mass term $\mu$ for
the two Higgs doublets, $\hhu$ and $\hhd$, which can neither vanish nor
be naturally large ($\sim M_{GUT}$ or $\sim M_{Pl}$) for phenomenological
reasons. The lack of an explanation of $\mu \approx M_{\susy}$ constitutes
the $\mu$-problem of the MSSM~\cite{Kim:1983dt}. 
The addition of a singlet chiral superfield, $\hs$, to the
particle content of the MSSM provides an elegant solution to the
$\mu$-problem: the scalar component of $\hs$ obtains a vacuum expectation
value (vev) of the
right order, dynamically generating the mass term $\mu$. 

Another challenge to these models is that the non-observation of the Higgs boson at 
LEP-II requires large soft-supersymmetry-breaking mass 
parameters to raise the mass of the lightest CP-even Higgs above 
the tree-level prediction, $m_h < m_Z$, in the MSSM. The discrepancy
between the large size of
these soft \susy\ breaking terms compared to their natural scale,
the electroweak scale, is known as the little-hierarchy 
problem. Here again, the scalar components of $\hs$, that mix with the 
neutral scalar components of $\hhu$ and $\hhd$, can alleviate the little fine-tuning
problem of the 
MSSM~\cite{BasteroGil:1999gu,BasteroGil:2000bw,Dermisek:2005ar} by
lifting the Higgs mass or allowing for new Higgs decay modes 
that weaken the LEP-II limits~\cite{Gunion:1996fb,Dobrescu:2000jt,Dermisek:2005ar}. The resulting model is the Next-to-Minimal Supersymmetric Standard Model
(NMSSM, see e.g.~\cite{Maniatis:2009re,Ellwanger:2009dp} for recent reviews), 
and its dark sector can differ considerably from that of the MSSM.
Much like in the Higgs sector, mixings with the fermionic component of
$\hs$, the singlino, result in an extended neutralino sector. 
The lightest neutralino can have
a sizeable singlino component and, if it is the LSP, the
expected signatures at colliders and the DM phenomenology could markedly 
differ from the minimal scenario. 

Interestingly, it has been pointed out that the positron excess observed 
by PAMELA can be explained by neutralino annihilation in the 
NMSSM~\cite{Bai:2009ka,Hooper:2009gm,Wang:2009rj}. The richer Higgs and
neutralino sectors can potentially accommodate the ingredients shown
in~\cite{Pospelov:2007mp,ArkaniHamed:2008qn,Nomura:2008ru} to result in
enhanced mostly leptonic fluxes. The fact that this scenario is motivated
independently from particle physics considerations as outlined above, makes
it even more appealing. It is this last point that we set forth to study
in this paper. We revisit in Section~\ref{sec:reviewsinglet} 
the region in the NMSSM parameter space that
allows for an explanation of the reported cosmic-ray anomalies in terms of
neutralino annihilations, and study whether the original motivation of
naturally addressing the little higgs and $\mu$-problem is preserved.
Our findings show that this is not generically possible without
accidental relations among the parameters, losing the naturalness motivation. 
In Section~\ref{sec:extension} we add another singlet superfield, $\hpsi$, 
to the dark sector, which suffices to
avoid reintroducing fine-tunings in the electroweak sector.
A study of a similar model, with vector-like dark matter, was presented in \cite{Nomura:2008ru} in 
the context of an axionic sector. Here we do not require a light 
singlet scalar to generate a Sommerfeld enhancement, as we take a light 
pseudoscalar to be sufficient. We study the dark matter sector in
Section~\ref{sec:dm} and we find somewhat different behavior in our scenario than the
one presented in~\cite{Nomura:2008ru} regarding the behavior of the extended NMSSM model.
Our results are summarized in Section~\ref{sec:concl}.

\section{Dark Matter Phenomenology in singlet extensions of the MSSM}
\label{sec:reviewsinglet} 

In supersymmetric extensions of the SM the masses of the up-type quarks
and down-type quarks are generated by the vevs of two Higgs SU(2)-doublets
$H_u$ and $H_d$. With this minimal field content in the Higgs sector, as
found in the MSSM, a dimension-full coupling $\mu \hat{H}_u \hat{H}_v$ 
must be appear in the superpotential. For phenomenological reasons the
parameter $\mu$ has to be of the order of the electroweak breaking scale, 
$v_{\mathrm{EW}}$, which is orders of magnitude below the natural value 
in the MSSM, $\mu\sim \Lambda$, where $\Lambda$ represents the 
ultra-violet (UV) cut-off of the 
theory (GUT or Planck scale). In the NMSSM (see
e.g.~\cite{Maniatis:2009re,Ellwanger:2009dp} for recent reviews and 
further references), the mass term $\mu$ is replaced by the
vev of a scalar field -- induced by the soft \susy\ breaking terms -- 
which has a Yukawa coupling to the Higgs doublets. 
The simplest scenario allows only scale-invariant and renormalizable terms 
in the superpotential:
\be
W = \lambda \hs \hhu \hhd + \frac{\kappa}{3} \hs^3,
\label{eq:siw}
\ee
where $\lambda$, $\kappa$ are dimensionless, and we have not included
couplings to the lepton and quark matter fields of the MSSM, 
which are not relevant here. 
Electroweak symmetry breaking generates vevs for the scalar components of 
both MSSM Higgs superfields, $\hhu$ and $\hhd$, as well as a vev
$\exv[S]$ for $\hs$ of the order of the weak scale.
In this way the NMSSM generates an effective $\mu$ paramater,
\beq
\mu = \lambda \exv[S],
\label{eq:muterm}
\eeq
of the order of the weak scale, $v_\mathrm{EW}\approx 174\GeV$,
which solves the $\mu$-problem of the MSSM. 
The scalar components of $\hs$ mix with the neutral scalar components
of $\hhu$ and $\hhd$, and this results (in the absence of explicit CP
violation) in three CP-even and two CP-odd neutral scalars. Mixing of
the fermionic components of $\hs$ with the neutral higgsinos and gauginos
leads to five neutralinos in the spectrum of the NMSSM. 
The tree level Higgs and neutralino mass matrices can be found 
in e.g.~\cite{Maniatis:2009re,Ellwanger:2009dp}, 
or deduced from the expressions
in Section~\ref{sec:extension} by taking the limit $\xi,\:A_\xi
\rightarrow 0$. 
 
As mentioned above, the richer Higgs sector in the NMSSM 
allows for a resolution of the little fine-tuning problem of the 
MSSM~\cite{Gunion:1996fb,Ellwanger:1999ji,Dobrescu:2000jt,Dermisek:2005ar}. 
In addition, the appearance 
of new Higgs decay modes, together with a possible singlino component,
can result in a markedly different phenomenology for the light neutralino, which
is a candidate for DM as in the MSSM.
Interestingly, it has been pointed out by several groups that an explanation of the anomalous
cosmic-ray fluxes reported by PAMELA in terms of NMSSM neutralino 
annihilations is possible.

In the scenario envisaged in~\cite{Bai:2009ka},
the LSP is a neutralino of the bino-type with Higgsino mixings. This
allows for a large annihilation cross-section into the lightest CP-even
plus CP-odd scalars, $\chi_1^0+\chi_1^0 \rightarrow h_1 + a_1$. 
As explained below,
it is technically natural for a CP-odd scalar in the NMSSM to be light, and
this restricts the subsequent decays to $a_1 \rightarrow \mu^+ \mu^-$.
Hence, a flux of energetic positrons is obtained, while kinematics prevents
the production of anti-protons, in agreement with observations. The 
positron fluxes can only be brought in agreement with the data, if
the mass of the heavier CP-odd scalar, $a_2$, is dialed to be
$m_{a_2} \approx 2 m_{\chi_1^0}$. Then, neutralino annihilation proceeds
through the resonant $a_2$ state, which would also increase the 
intensity of monochromatic $\gamma$-ray lines~\cite{Ferrer:2006hy}. 
This scenario exemplifies the possibilities allowed by the larger
Higgs sector of the NMSSM, but the careful choice of parameters
required for the resonant annihilations to occur opposes the initial motivation
to go beyond the MSSM.

These models introduce further theoretical questions.
For instance, the annihilation cross-section is more than two orders
of magnitude larger {\it at all times}, 
so that a non-thermal production mechanism has to be invoked to match 
the observed DM density. 
Since a non-thermal neutralino in the MSSM is also capable of
explaining the PAMELA signal~\cite{Grajek:2008pg}, there would be 
little motivation to go beyond the MSSM. 
Furthermore, in the scenario presented in \cite{Bai:2009ka}, the loop corrections to the mass of $a_1$ are not suppressed by any small parameter, so a fine-tuning is required to achieve the sub-GeV mass in their parameter set that primarily produces muons in dark matter annihilation. The neutralino is also required to be less than the top mass, which may present a difficulty in fitting the shape of the PAMELA spectrum.

A different possibility, considered in~\cite{Hooper:2009gm}, is that
the neutralino could have a sizeable singlino component in the NMSSM. 
Annihilations would still proceed to the lightest scalars $h_1+a_1$, both
of which would now be mostly singlet-like, not just $a_1$. 
Due to its very small couplings to all quarks, leptons and gauge bosons,
a singlet-like $h_1$ is not bound by the LEP-II limit, $m_{H_{SM}}\gtrsim
114$ GeV, and could in fact be much lighter. Such a light scalar would
naturally enhance the rate of neutralino annihilations in the Galaxy 
through the Sommerfeld enhancement, without the need of resonant annihilations
and a non-standard cosmological evolution. Hence, in this setting 
all the ingredients
of the DM scenarios engineered to reproduce the PAMELA 
signal~\cite{Pospelov:2007mp,ArkaniHamed:2008qn,Nomura:2008ru} would be
realized in a framework motivated {\it independently} from considerations
of naturalness of the electroweak interactions. One should still make sure,
however, that this does not require particular choices of the
parameters to solve the $\mu$-problem, which was the main motivation
to go beyond the MSSM~\footnote{The mass of the next-to-lightest
Higgs scalar with SM-like couplings
can still be larger than in the MSSM, and this alleviates the
little fine-tuning problem~\cite{Ellwanger:1999ji}}.  

For a sufficient enhancement of the DM annihilation cross-section, the 
mass of the lightest CP-even scalar, $h_1$, can be estimated as
$m_{h_1} \lesssim \kappa^2 m_{\chi^0}/4 \pi$, where $\kappa$ is the
trilinear coupling in Eq.~(\ref{eq:siw}). The anomalous trend in the
positron fraction reported by PAMELA extends up to energies 
of $\sim 100$ GeV, which also sets the (minimum) mass of the neutralino.
For moderate values of $\kappa \lesssim 1$ the mass of $h_1$ should
be roughly $m_{h_1}\lesssim 10$ GeV (more details can be found 
in~\cite{Hooper:2009gm,ArkaniHamed:2008qn}). 
Since the decay of the lightest CP-odd scalar, $a_1$, should yield 
mostly charged leptons ($e^\pm$ or $\mu^\pm$), we demand $m_{a_1}
\lesssim 1$ GeV. We also require that the $\mu$ term is dynamically generated
as in Eq.~(\ref{eq:muterm}).

\begin{figure}[htpb]
\includegraphics[width=10cm]{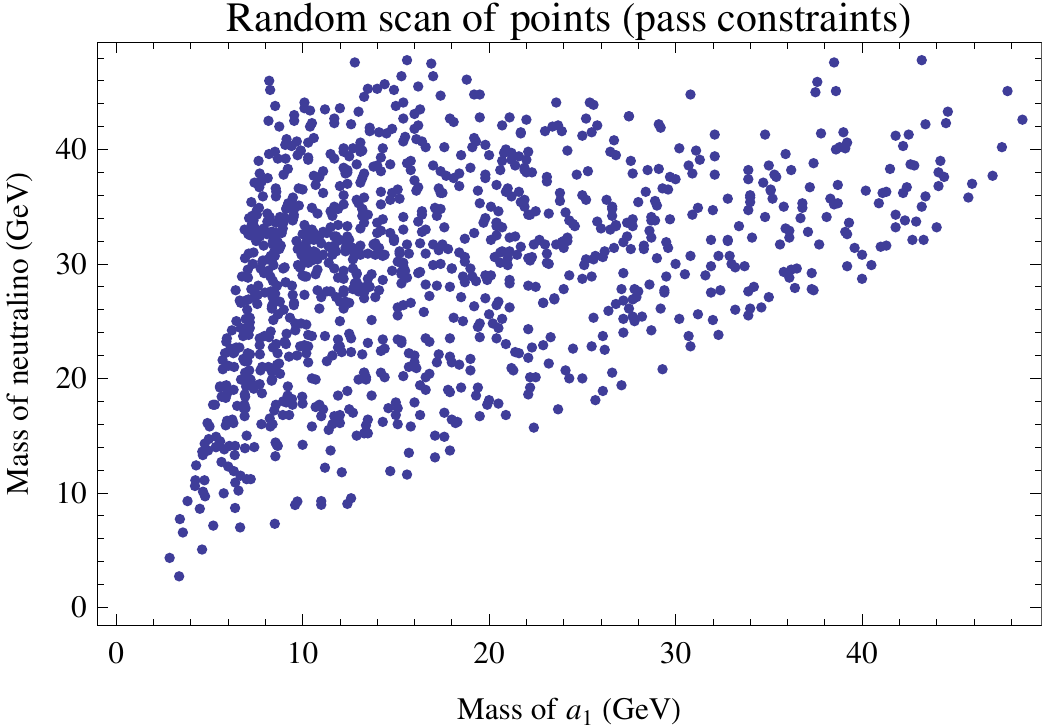}
\caption{Values of the mass of the light pseudoscalar and the neutralino for parameter sets that pass constraints for the scan described in the text.}
\label{fig:scan1}
\end{figure}

The parameter space of the NMSSM defined by Eq.~(\ref{eq:siw}) can be
studied with the package 
NMSSMTools \footnote{{\tt http://www.th.u-psud.fr/NMHDECAY/nmssmtools.html}},
which calls MicrOMEGAS\footnote{{\tt http://wwwlapp.in2p3.fr/lapth/micromegas}}
to calculate the relic density~\cite{Ellwanger:2005dv,Belanger:2008sj}.
Given a point in the NMSSM parameter space, the program suites check 
its viability against negative particle searches and rare-decay bounds from 
accelerators. 
We performed a number of scans, both over random points and on grids, in different regions of parameter space. For instance, in the region defined above, which is the region considered by \cite{Bai:2009ka}, we randomly selected one million points that had $\lambda<0.6$, with a flat distribution, and $\kappa<\lambda$. We fixed the soft terms $A_\lambda$ and $A_\kappa$ to have an exponential distribution peaked at zero, with a width of 10 GeV, to attempt to take advantage of a $U(1)_R$ symmetry that makes a pseudoscalar light, as in \cite{Bai:2009ka}. We also fixed $\tan\beta$ between 2 and 10, and $\mu$ between 100 and 240 GeV, and we randomly assigned all free signs to be positive or negative. The results demonstrate the difficulty in finding valid models. Approximately 0.2\% of the parameters passed all present experimental constraints. Of these, none had a pseudoscalar below 1 GeV. Additionally, the neutralino had a mass of order the light scalars, $\mathcal{O}(10\GeV)$ rather than the required 100 GeV or greater. The spectrum of points that passed LEP and other constraints is shown in Figure \ref{fig:scan1}.

On the other hand, non-scale invariant terms in the superpotential were
allowed in~\cite{Hooper:2009gm}. Including supersymmetric mass and
tadpole terms, $\mu \hhu \hhd + \frac{1}{2}\mu' \hs^2 + \xi_F \hs$, it
is possible to viable points in this extended parameter 
space~\cite{Wang:2009rj}. Even though, some of these
terms are non-vanishing in various scenarios, they contradict the theoretical
motivation for the NMSSM that the $\mu$-term should be dynamically generated
as in Eq.~(\ref{eq:muterm}). We cannot exclude the possibility 
that a more throughout scan might reveal a valid region that meets 
our stricter requirements, but the general considerations below should
illustrate the difficulties for a generic natural extension of the MSSM.

Light CP-odd scalars can easily appear in the NMSSM in the form of 
(pseudo-)Nambu-Goldstone bosons associated to approximate global symmetries.
Both, in the R-symmetry limit, $A_\lambda, A_\kappa \rightarrow 0$,  
considered in~\cite{Bai:2009ka,Hooper:2009gm,Wang:2009rj}, 
and in the Peccei-Quinn (PQ) limit, $\kappa \rightarrow 0$, on which
we focus below, axion-like
particles would be naturally present in the spectrum. Concrete scenarios
exist (as reviewed in e.g.~\cite{Maniatis:2009re,Ellwanger:2009dp}),
in which this symmetries are explicitly broken, yet the axion remains light. The anti-proton constraints can, thus, be naturally met in the NMSSM.

There is no symmetry, however, that prevents the mass of the CP-even
scalar, $h_1$, from growing above the scale required for the Sommerfeld
enhancement to be effective. In fact, as shown in Section~\ref{sec:extension},
requiring that the $\mu$-parameter
is generated as in Eq.~(\ref{eq:muterm}) and that the lightest
neutralino be heavier than $\sim 100$ GeV, cannot be simultaneously
achieved while keeping $m_{h_1} \lesssim 10$ GeV. 
If we allow for non-scale invariant terms in the superpotential as 
in~\cite{Hooper:2009gm,Wang:2009rj}, the lightest scalar is no longer linked
to the electroweak-scale $\mu$-term and can be lighter. However, this is 
not a natural outcome of singlet extensions of the MSSM, and requires
the tuning of the parameters.

The main reason to include a light CP-even scalar is to enhance the annihilation
cross-section in the low velocity limit, so that the positron fluxes
from annihilations in the Galaxy can be large enough, while obtaining the
right relic density from thermal freeze-out at the time of decoupling. 
As outlined in~\cite{Hisano:2004ds,Cirelli:2007xd,ArkaniHamed:2008qn}, 
such a light particle mediates long range forces that generate the
non-perturbative Sommerfeld enhancement. A light vector particle also results
in long-range forces, and this was in fact the first scenario
in which the importance of this effect for 
DM was studied~\cite{Hisano:2004ds}. Other possibilities, such as
CP-odd mediator particles, have been dismissed, since they do not give rise
to spin-independent long range interactions {\it at tree level}. However,
since pseudo-scalars do give rise to long-range forces at the
one loop level~\cite{Ferrer:1998ue}, this conclusion should not necessarily
apply in the non-perturbative regime. Indeed, as shown
in~\cite{Bedaque:2009ri}, (pseudo-)Nambu-Goldstone bosons can
cause a Sommerfeld enhancement. As discussed above, we do have light
pseudo-scalars in our spectrum, and we might ask whether lifting the limit
on the mass of the $h_1$, allows a DM explanation to the PAMELA observations,
while abiding to Eq.~(\ref{eq:muterm}).

Unfortunately, we find that this is not sufficient. In order to generate a splitting between the scalar and fermion components of the $\hs$ superfield, in general we need to increase the mixing with the component of $\hhu$ and $\hhd$ by increasing $\lambda$. However, this also increase the magnitude of radiative corrections to the mass of the light scalars, requiring a high-degree of fine-tuning to maintain a sub-GeV pseudoscalar mass. One might hope to escape this problem by introducing additional superpotential terms, as in \cite{Hooper:2009gm}, however a subsequent study of the parameter space by \cite{Wang:2009rj} found that this scenario requires $\lambda\sim 10^{-3}$, a small coupling with no clear origin.

The previous discussion shows the difficulties in constructing a DM sector
that allows an explanation of the rising positron fraction, and that
is motivated by naturalness considerations of the electroweak interactions.
A possible solution, with only scale invariant terms, was
put forward in~\cite{Nomura:2008ru}, where it was noted that 
an additional singlet could be added to the PQ-limit of the NMSSM.
In the following we consider an scenario in the spirit 
of~\cite{Nomura:2008ru}, and study the restrictions imposed by
Eq.~(\ref{eq:muterm}). In addition, we note that a similar superpotential was proposed
in~\cite{Cerdeno:2008ep} to extend the NMSSM by a right-handed neutrino superfield,
which serves as a suitable thermal dark matter candidate.
As the motivation in~\cite{Cerdeno:2008ep} is not to explain observations made by PAMELA, the
authors examine a different region of parameter space and an additional superfield coupling
that is not relevant for our purposes.

\section{Adding a Dark Matter Singlet}
\label{sec:extension}
Following from the above consideration of the NMSSM, we argue that a supersymmetric model that is not fine-tuned and contains leptophilic dark matter annihilation requires extending the MSSM by more than one field. Here we will consider the simplest case: the MSSM with an additional two superfields: a gauge singlet $\hs$, as in the NMSSM, and an additional singlet $\hpsi$ that will serve as dark matter. In our arrangement, the $\hs$ and $\hpsi$ together form a dark sector that is moderately secluded from the MSSM fields. While at first glance this theory may appear to be overly general, we will show below that observations and naturalness significantly restrict the allowed couplings.


To naturally generate a light pseudoscalar, we will take the $U(1)_{PQ}$ limit of the NMSSM, but will include an explicit symmetry breaking term $\kappa$. We take this limit both because it generates a light pseudoscalar with a controllable mass, and because that small mass is more stable against radiative corrections than the one provided by $U(1)_R$ symmetry. We assume the $\hpsi$ superfield has a $\mathbb{Z}_2$ symmetry which will guarantee a stability.

The superpotential of this theory is
\beq
W = \frac{\xi}{2} \hs \hpsi^2 + \lambda \hs \hhu \hhd + \frac{\kappa}{3} \hs^3,
\eeq
where $\hs$ is the singlet of the NMSSM,  and $\hhu$ and $\hhd$ are electroweak Higgs doublets, as in the MSSM. The new superfield $\hpsi$ consists of a fermionic component $\psi$, which will be a stable dark matter candidate, and complex scalar $\phi$. We have not included the other matter fields of the MSSM, which are not relevant here. The corresponding scalar soft terms are
\beq
V_{\mathrm{soft}} = \frac{\xi}{2} A_\xi S \phi^2-\lambda A_\lambda S H_u^0 H_d^0+\frac{\kappa}{3} A_\kappa S^3+\mathrm{h.c.},
\eeq
where we have dropped electrically charged components. Since $\kappa$ is a small breaking of the $U(1)_{PQ}$ symmetry, we will take $\kappa \ll 1$. The typical sizes of the other parameters are discussed below. 


As in the NMSSM, we require that the effective $\mu$ parameter given in (\ref{eq:muterm}) is of order the electroweak scale. Electroweak symmetry breaking does not generate a VEV for the scalar partner of the dark matter, $\phi$. The consistency of this minimum can be checked by examining the eigenvalues of the Hessian for the scalar potential. In this paper we will not study the global structure of minima.

\section{Mass spectrum}

Having defined the theory, we will turn our attention to the mass spectrum of the theory in the presence of the additional states. We will denote the real scalar and pseudoscalar components of the $\phi$ and $S$ fields with the subscripts $s$ and $a$, respectively, and use the usual notation for the Higgs VEVs,
\beq
\exv[H_u^0]=v_u, \exv[H_d^0]=v_d,
\eeq
where $v_u^2+v_d^2=v_\mathrm{EW}^2$. In reference to the MSSM, it will be convenient to define an effective $B$ parameter,
\beq
B=A_\lambda+\kappa\exv[S]=A_\lambda+\kappa\frac{\mu}{\lambda}.
\eeq

The CP-even mass matrix, in the basis $(H_d,H_u,S_s,\phi_s)$, is given by
\beq
M_{RS}^2=
\left( \begin{array}{cccc}
g^2v_d^2+B\mu\frac{v_u}{v_d} & v_u v_d(2\lambda^2-g^2)-B\mu & 2\lambda \mu v_d-v_u(\lambda B+\kappa \mu) & 0\\
 & g^2v_u^2+B\mu\frac{v_d}{v_u} & 2\lambda\mu v_u-v_d(\lambda B+\kappa\mu) & 0 \\
 & & v_uv_d(\frac{\lambda^2 B}{\mu}-\kappa\lambda)+\kappa A_\kappa\frac{\mu}{\lambda} & 0 \\
 & & & \lambda\xi v_u v_d+\xi A_\xi \frac{\mu}{\lambda}+\kappa\xi\frac{\mu^2}{\lambda^2}+\xi^2\frac{\mu^2}{\lambda^2}
\end{array} \right)
\eeq
where we drop terms of $\mathcal{O}(\kappa^2)$.

The Higgs sector contains a neutral massless Goldstone mode, $G$, that becomes the longitudinal mode of the $Z$ boson. This can be rotated away, leaving a single physical Higgs pseudoscalar from the MSSM, $A$. After performing this transformation, the CP-odd mass matrix, in the basis $(A,S_a,\phi_a)$, is given by
\beq
M_{IS}^2=
\left( \begin{array}{ccc}
\frac{B\mu(v_d\cos\beta+v_u\sin\beta)^2}{v_u v_d} & (B\lambda-3\kappa\mu)(v_d\cos\beta+v_u\sin\beta) & 0 \\
 & \frac{\lambda v_u v_d}{\mu}(B\lambda+3\kappa\mu)-3\frac{\kappa}{\lambda} A_\kappa \mu & 0 \\
 & & \frac{\xi}{\lambda^2}(v_uv_d\lambda^3-\mu(\lambda A_\xi+\kappa\mu-\xi\mu))
\end{array} \right)
\eeq

The neutralino mass matrix is extended from the MSSM by the fermionic components of the superfields $S$ and $\Psi$. In the basis $(\tilde{B},\tilde{W}^0,\tilde{H_d},\tilde{H_u},\tilde{S},\psi)$, it is given by
\beq
M_F = 
\left( \begin{array}{cccccc}
M_1 & 0 & -\frac{g_1 v_d}{\sqrt{2}} & \frac{g_1 v_u}{\sqrt{2}} & 0 & 0 \\
 & M_2 & \frac{g_2 v_d}{\sqrt{2}} & -\frac{g_2 v_u}{\sqrt{2}} & 0 & 0 \\
 & & 0 & -\mu & -\lambda v_u & 0 \\
 & & & 0 & -\lambda v_d & 0 \\
 & & & & 2 \frac{\kappa}{\lambda} \mu & 0 \\
 & & & & & \frac{\xi}{\lambda} \mu
\end{array} \right)
\eeq

We will label the resulting scalar and pseudoscalar mass eigenstates with $h_i$ and $a_i$ respectively, ordered from lightest to heaviest. Similarly, we will label the neutralinos with $\tilde{N}_i$. To avoid confusion, we will continue to label the states of $\Psi$ with $\psi$ and $\phi$, since these states are the same as their flavor states.

We will be interested in the region of parameter space where $\lambda$ is small, but not unnaturally so. In the PQ limit, which we consider, $\kappa$ is expected to be very small, but $\lambda$ is not. In this case, the lightest scalar and pseudoscalar, $h_1$ and $a_1$, are composed primarily of singlet $S$ states. The Higgs with significant couplings to the Standard Model are $h_2$, $h_3$ and $a_2$. The lightest neutralino, $\tilde{N}_1$, is composed primarily of singlino $\tilde{S}$. Since the lightest states are all only slightly mixed with other Higgs, their coupling are very nearly those of the $\hs$ flavor states.

The behavior of dark matter in this theory primarily depends on the value of the masses of the light particles. In the region of parameter space we will look at below, the Sommerfeld enhancement is generated primarily by $a_1$. The lightness of this particle also kinematically enforces the creation of light leptons rather than baryons in dark matter annihilation. To leading order in $\kappa$ and $\lambda$, the tree-level mass of this particle is
\beq
m_{a_1}^2 = \frac{9}{2}v_\mathrm{EW}^2\kappa\lambda\sin 2\beta.
\label{eq:ma2} 
\eeq

\begin{figure}[htpb]
\includegraphics[width=7.5cm]{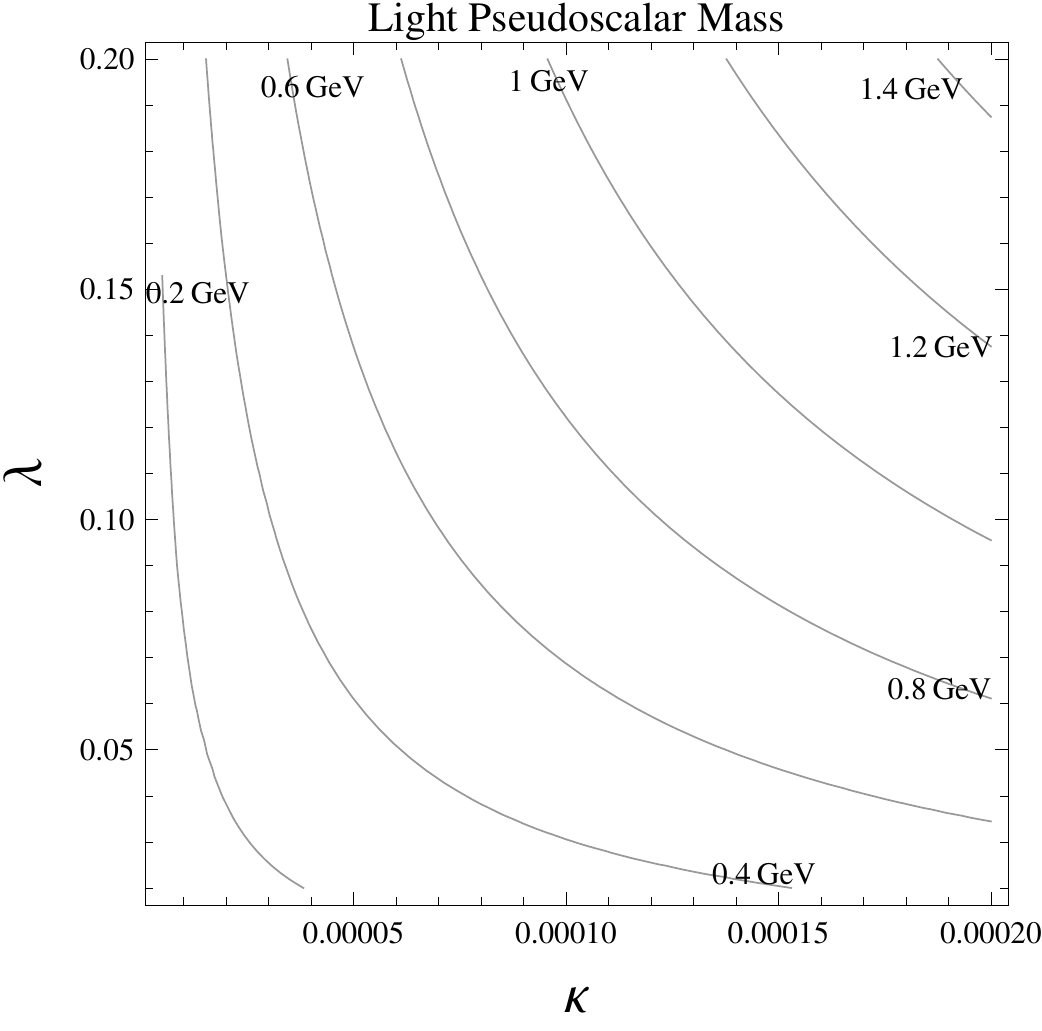}
\includegraphics[width=7.5cm]{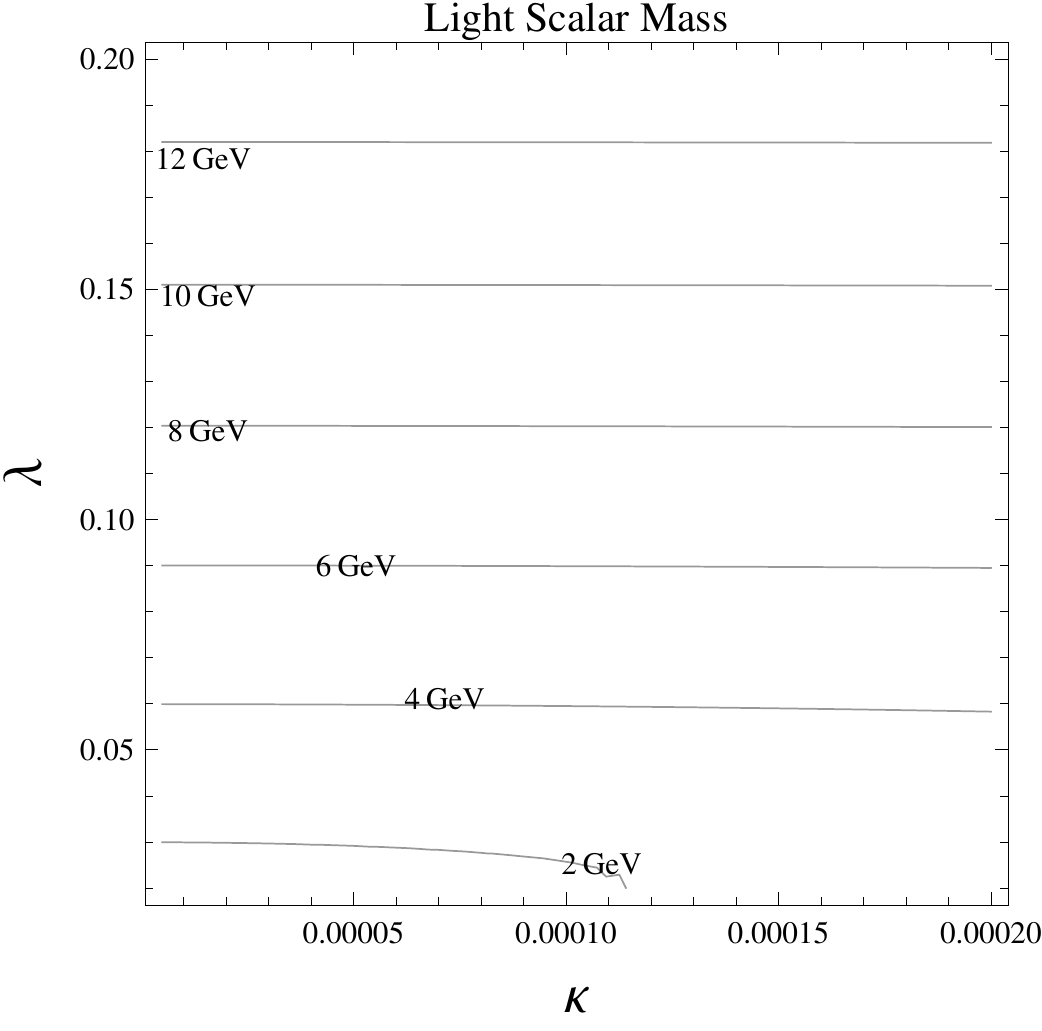}
\caption{Contour plots of the tree-level values of $m_{a_1}^2$ and $m_{h_1}^2$ as a function of $\kappa$ and $\lambda$ for the case of $\mu=130\GeV$ }
\label{fig:ahcontour}
\end{figure}

The lightest scalar, $h_1$, has a leading mass $m_{h_1}$ that is independent of $\kappa$ and so is generically unsuppressed relative to $m_{a_1}$. The more complicated mixing of the $3\times 3$ CP-even mass submatrix does not lend itself to compact expression for this mass. To leading order in $\lambda$, the mass is given by
\beq
m_{h_1}^2=\frac{-\lambda^2 \sec^2 2\beta}{2g^2 A_\lambda \mu} 
&&\left( 8 A_\lambda \mu^3 - 8A_\lambda^2\mu^2\sin 2\beta
+2A_\lambda\mu(A_\lambda^2-2g^2v^2)\sin^22\beta \right. \nonumber \\
&&+ \left.  g^2v^2(A_\lambda^2+4\mu^2)\sin^32\beta \right) .
\label{eq:mh1}
\eeq
The hierarchy of scalar and pseudoscalar masses is illustrated for typical values in Figure \ref{fig:ahcontour}.

\begin{figure}[htpb]
\includegraphics[width=10cm]{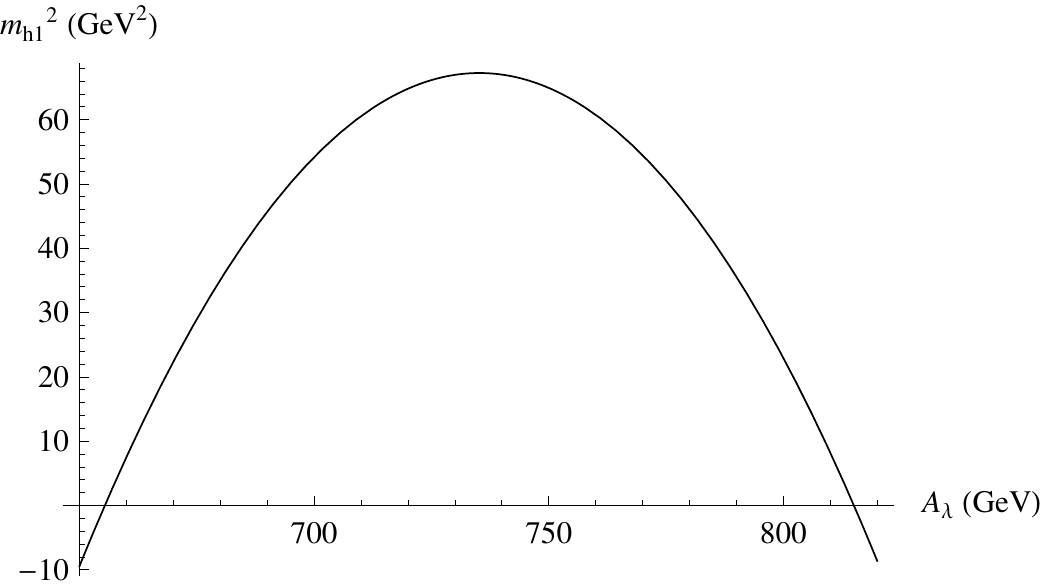}
\caption{Values for the tree-level square mass of the light scalar, $m_{h_1}^2$, where $\mu = v_\mathrm{EW}$, $\lambda=0.1$, $\tan\beta=4$ and $A_\lambda=500\GeV$.}
\label{fig:ms2}
\end{figure}

Positivity requirements for the mass place restrictions on the allowed values of the other coupling constant and soft terms. This relationship further implies the existence of a maximal value for $m_{h_1}$ once the dimensionless coupling constants are fixed. An example of this for typical values is given in Figure \ref{fig:ms2}. $A_\lambda$ is of the order of the typical soft breaking scale. $A_\kappa$ is less relevant, as its contribution is suppressed by $\kappa$.

The scalar and pseudoscalar masses are shifted by loop corrections from the tree-level values given above. These corrections are suppressed by additional powers of $\lambda$ and $\kappa$, plus loop factors. For $\lambda \lesssim 0.1$, these corrections are sufficiently small that they do not affect the results we present. In particular, the light pseudoscalar receives mass corrections that are less than a GeV.
Note that this is true in the PQ-limit with small
$\kappa$, but requires an unnaturally small value of $\lambda$ 
(which is not related to any enhanced symmetry) in the
regions considered in~\cite{Bai:2009ka,Hooper:2009gm,Wang:2009rj}.

The supersymmetric partner to these light scalars, $\tilde{N}_1$, typically has a mass of a few GeV that is intermediate between $h_1$ and $a_1$. To leading order in $\lambda$, $g_1$ and $g_2$, an approximation good to 10\% or better, this particle has a tree-level mass
\beq
m_{\tilde{S}}=\frac{\lambda^2 v^2\left( 4M_1M_2\mu\sin 2\beta-2(g_2^2M_1+g_1^2M_2)v^2\cos^22\beta\right)}{4M_1M_2\mu^2}.
\eeq

Finally, there is the mass of the dark matter itself. The fermion $\psi$ is stable and has a mass that is fixed by $\exv[S]$ and the coupling $\xi$,
\beq
m_\psi = \xi\frac{\mu}{\lambda}.
\label{eq:massdm}
\eeq
Note that if $\mu$ is near the electroweak scale, then $\mu/\lambda$ is $\mathcal{O}(\mathrm{TeV})$ for the relevant values of $\lambda$. We then get a dark matter mass of $\mathcal{O}(100\GeV)$ or above for $\xi\gtrsim 0.1$.

\section{Dark Matter}
\label{sec:dm}

Within the model described above, we identify several possible dark matter scenarios. First, we assume that supersymmetry breaking is gauge-mediated, so the true LSP of the theory is the gravitino. The stability of $\psi$ is protected by an unbroken $\mathbb{Z}_2$ symmetry, which makes it a suitable heavy dark matter candidate.  The $\phi_s$ and $\phi_a$ have mass that is typically somewhat larger, so can decay through the channel $\phi_{s,a}\rightarrow \psi \tilde{N}_1$. The $\tilde{N}_1$, the NLSP of the theory, will then decay to a gravitino and a singlet scalar.

Even if $m_\phi < m_\psi+m_{\tilde{N}_1}$, the $\phi$ can still decay to a gravitino and a $\psi$, either in tree-level processes with a virtual $\tilde{N}_1$, or in 1-loop processes with an $h_1$ or an $a_1$ in the loop. To avoid problems in the early universe, we assume either that $\phi$ is sufficiently massive to decay to $\psi$ and $\tilde{N}_1$ on-shell, or that the suppressed processes allow the $\phi$ to decay prior to the BBN.

If the $\phi$ are long-lived due to suppressed decay processes, a careful consideration of the early universe is required. If a significant number of $\phi$ particles remain after freeze-out, a late decay to gravitinos after nucleosynthesis will ruin the successful predictions of early universe cosmology. However, the $\phi$ has additional self-annihilation channels generated by 3- and 4-point scalar interactions that tend to raise its cross-section significantly, and their relic density becomes irrelevant. If the $\phi$ mass is nearly degenerate with the $\psi$ mass, then there may be additional cross-annihilation channels that will shift the $\psi$ relic density from the values given below. This offers the interesting possibility of breaking the relationship of $m_\psi$ and the relic density, both of which are set by $\xi$.

There is also the possibility that supersymmetry breaking is gravity-mediated. In this case, the singlino is the true LSP and absolutely stable. This is problematic in the model presented above, as the dominant channel for singlino self annihilation, to Higgs through an s-channel, scales as $\kappa^2$, and so $\tilde{N}_1$ is overproduced thermally in the early universe. However, in a slightly expanded version of the model, in which we allow dimensional superpotential terms, we can arrange for a $U(1)_{PQ}$ symmetry with non-infinitesimal $\kappa$. A version of this model was described in \cite{Hooper:2009gm}. Here the self-coupling can be much larger, and we can arrange for a two-component dark matter system that is a mixture of the heavy $\psi$ and the light $\tilde{N}_1$. Since this scenario requires an extended parameter set, we will not consider it further here.




We will make the new fields moderately secluded through a small, but natural, value of the coupling $\lambda\sim 0.1$. Above this value, the radiative corrections to the scalar masses shifts their values from the tree-level significantly. Below this value, the annihilation cross-section is suppressed, and the universe tends to overclose. After $\lambda$ and $\mu$ are fixed, the coupling $\xi$ is still free to fix both the dark matter mass and its annihilation cross-section. We will show below that $\xi\sim 0.4$ is an appropriate value. Our couplings are arranged in the loose hierarchy 
\beq
\kappa \ll \lambda < \xi.
\eeq

\subsection {Relic Density}

The relic dark matter is thermally produced in this scenario. While the coupling of the dark matter to the standard model fields is small, the coupling $\xi$ of the dark matter particles to fields of $\hs$ is essentially unconstrained. The value of this parameter fixes the annihilation cross-section, and it is notable that the same values that produce dark matter mass of $\mathcal{O}(100\GeV)$ in (\ref{eq:massdm}) also yield the observed relic density.

A number of annihilation channels are potentially relevant. There is an s-channel process mediated by $h_1$ or $a_1$ with a two-particle final state consisting of a combination of scalars $h_1$, $a_1$, $h_2$ or $h_3$. Higgsino production in this channel is possible, but suppressed by $\kappa$ if it is allowed kinematically.

\begin{figure}[htpb]
(a)
\includegraphics[width=3.8cm]{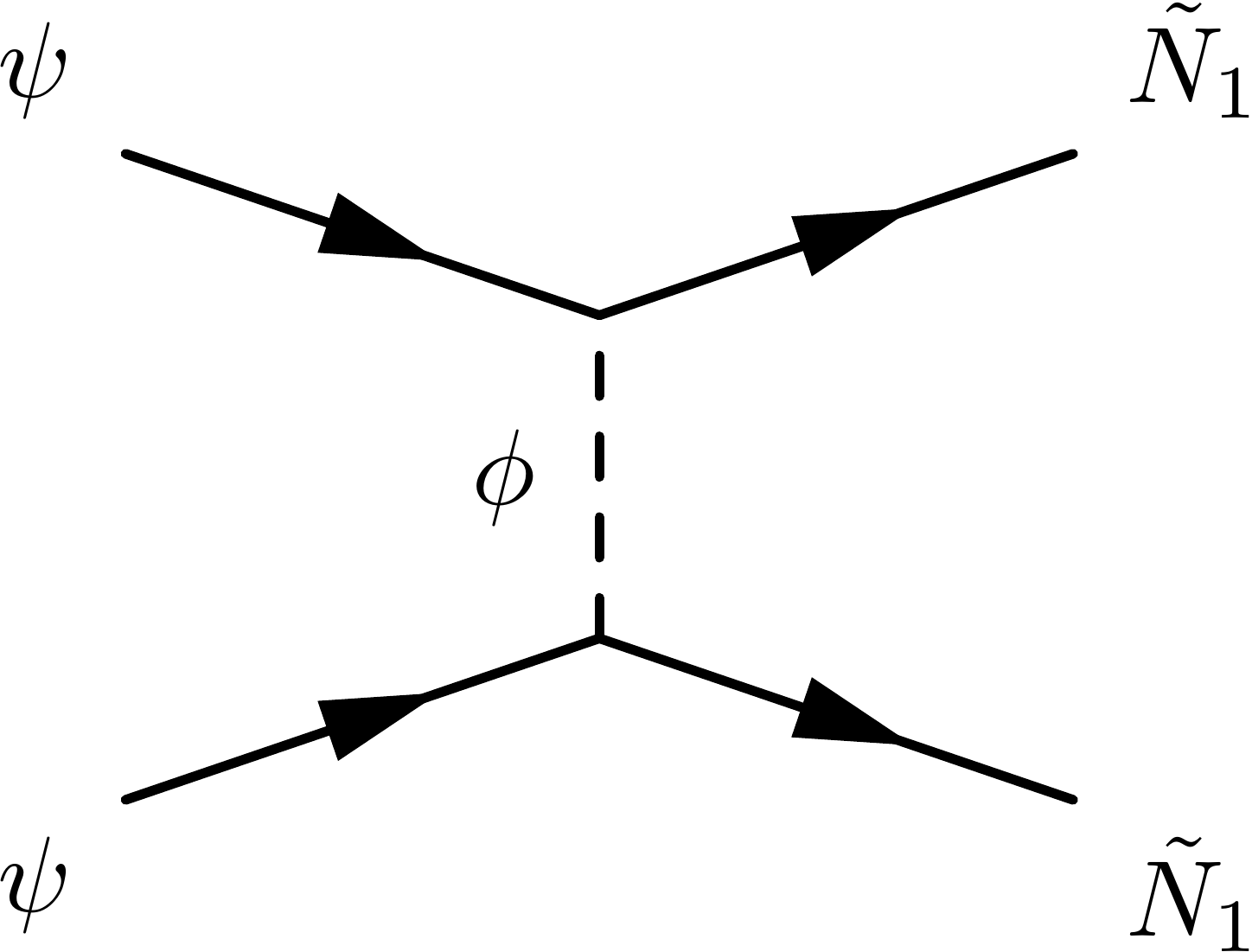}
\hspace{1cm}
(b)
\includegraphics[width=3.8cm]{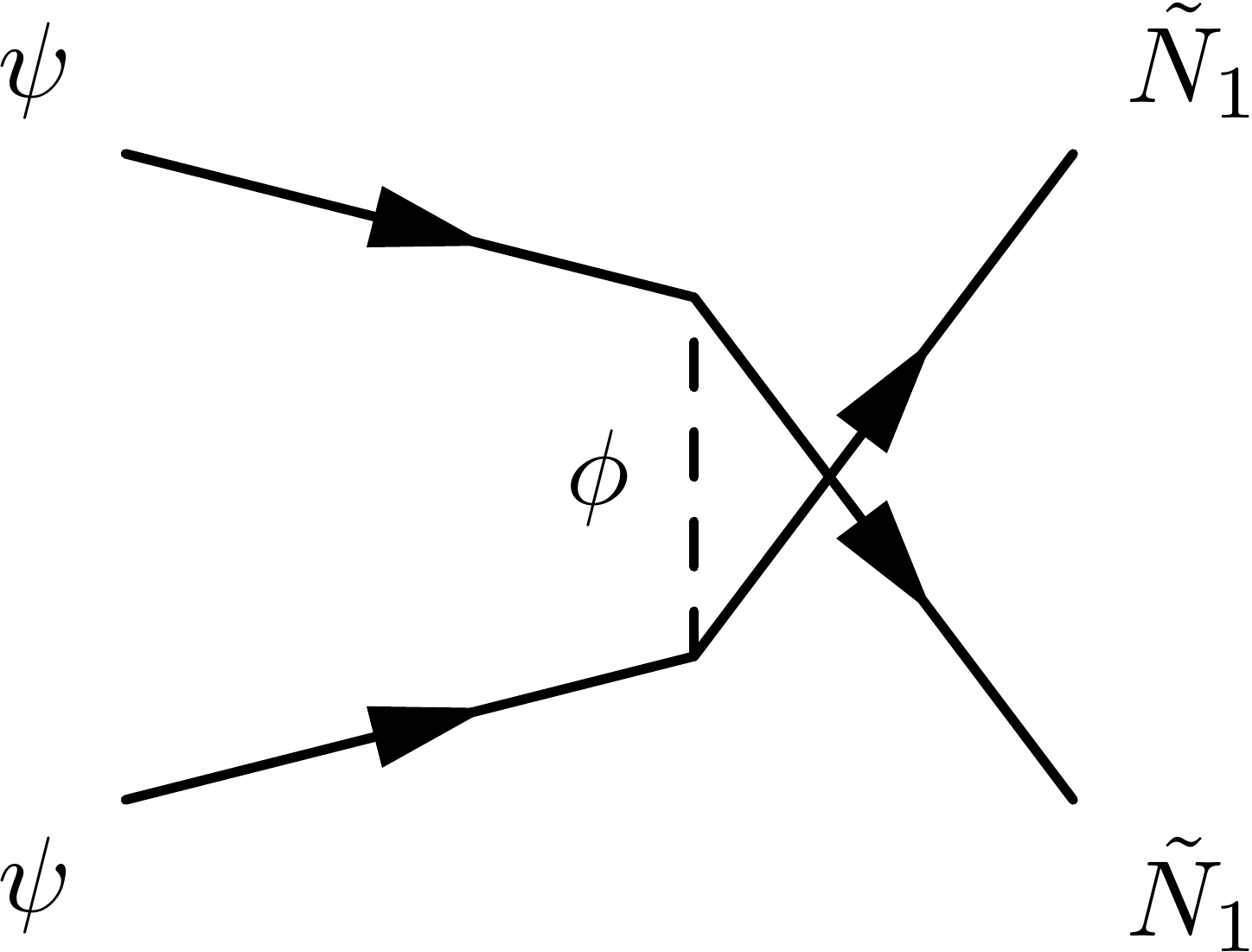}

\vspace{1cm}
(c)
\includegraphics[width=5cm]{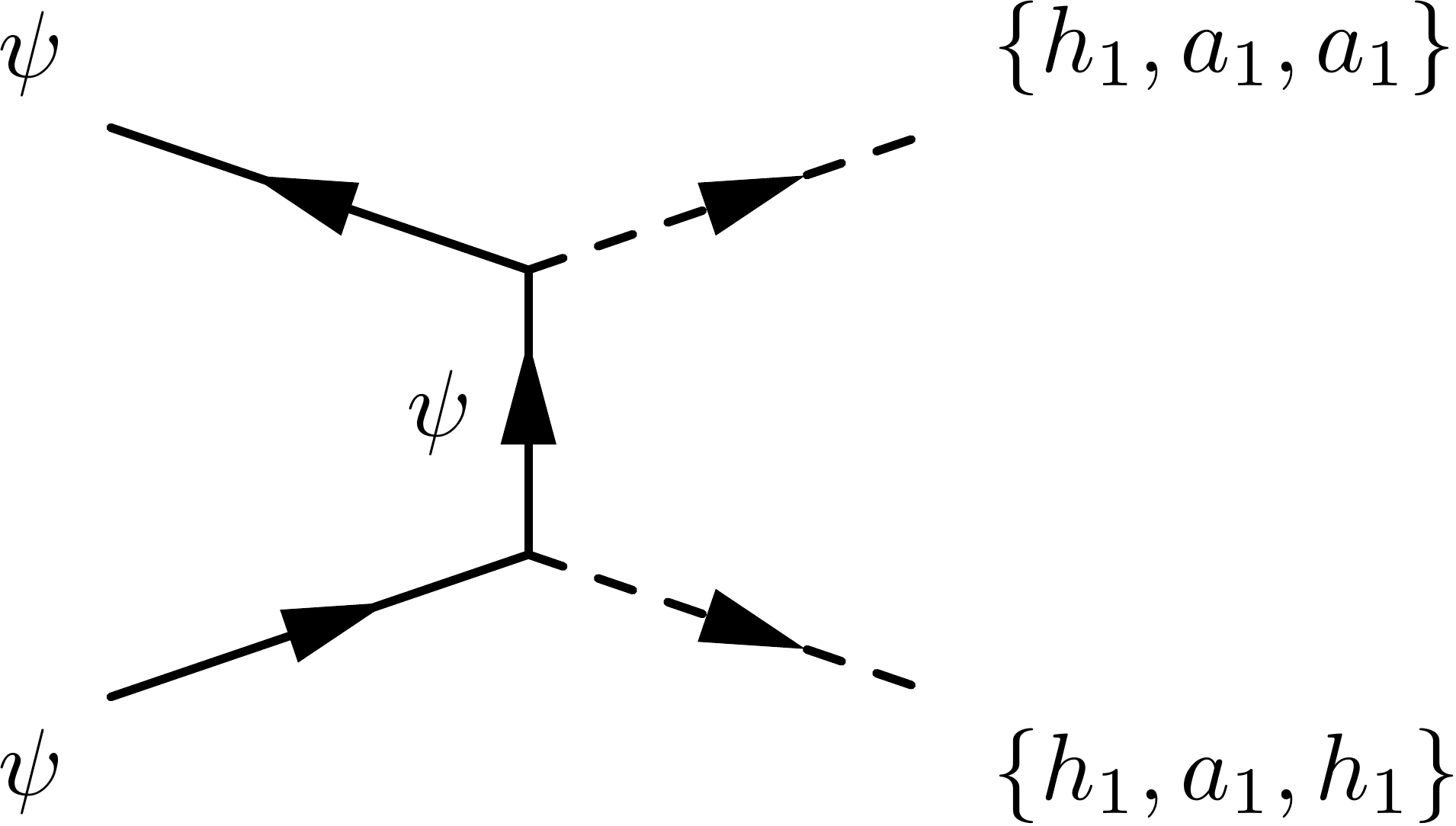}
\hspace{1cm}
(d)
\includegraphics[width=5cm]{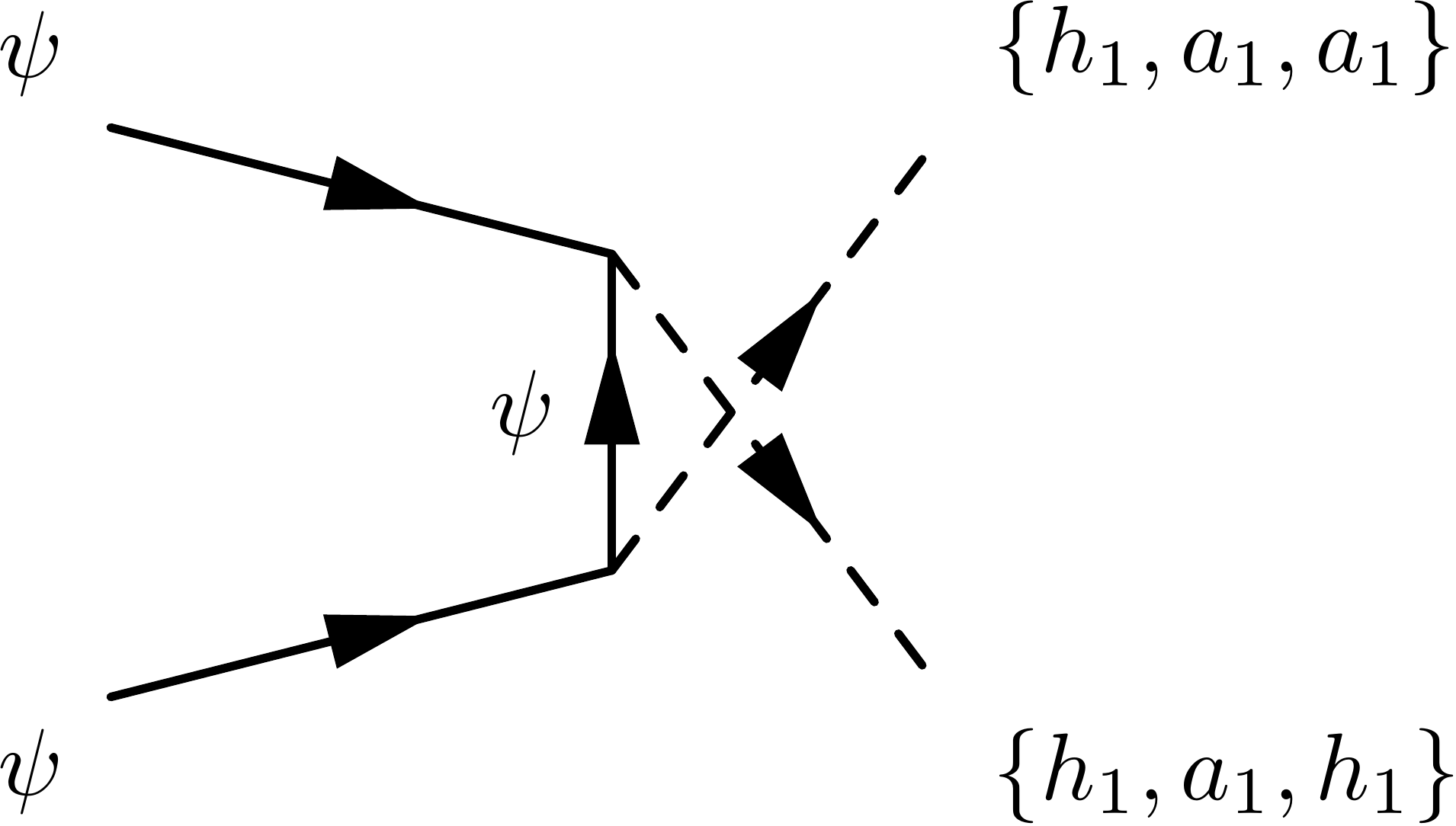}

\vspace{1cm}
(e)
\includegraphics[width=7.5cm]{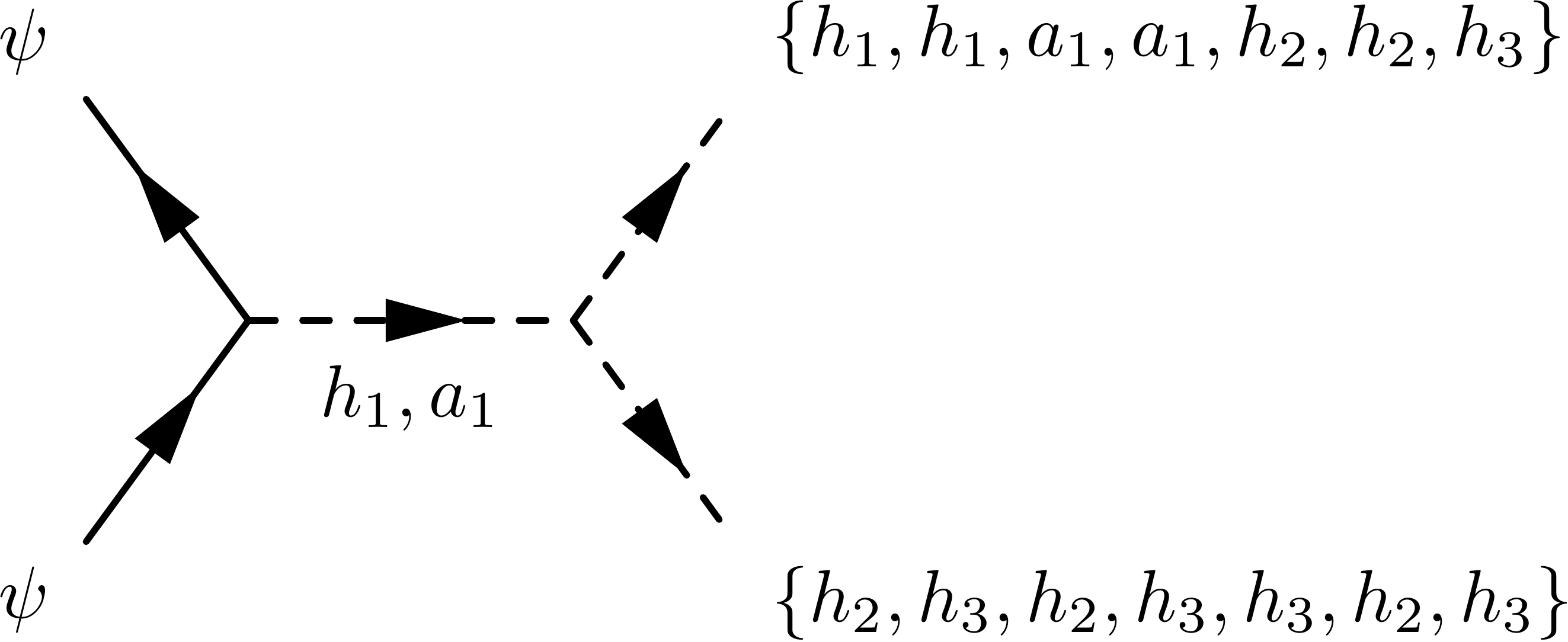}
\caption{Processes that contribute to the annihilation of $\psi$ which are unsuppressed by $\kappa$. }
\label{fig:psipsidiag}
\end{figure}

There are two relevant t- and u-channel processes. The first is mediated by $\phi$ and results in two $\tilde{N}_1$ particles in the final state. The second is mediated by $\psi$ and results in an $h_1$ and $a_1$. The processes that are not $\lambda$-suppressed are shown in Figure \ref{fig:psipsidiag}.

When comparing the cross-sections of the various channels, one might expect that the Higgs production is suppressed by a large amount, as these matrix elements appear to be proportional to $\lambda^2$. However, some channels are enhanced by $\exv[S]$, which is proportional to $\lambda^{-1}$, and they are only
suppressed by one power of $\lambda$. Since we do not have $\xi \gg \lambda$, we can not \emph{a priori} ignore Higgs production during $\psi$ annihilation.

\begin{figure}[htpb]
\includegraphics[width=8cm]{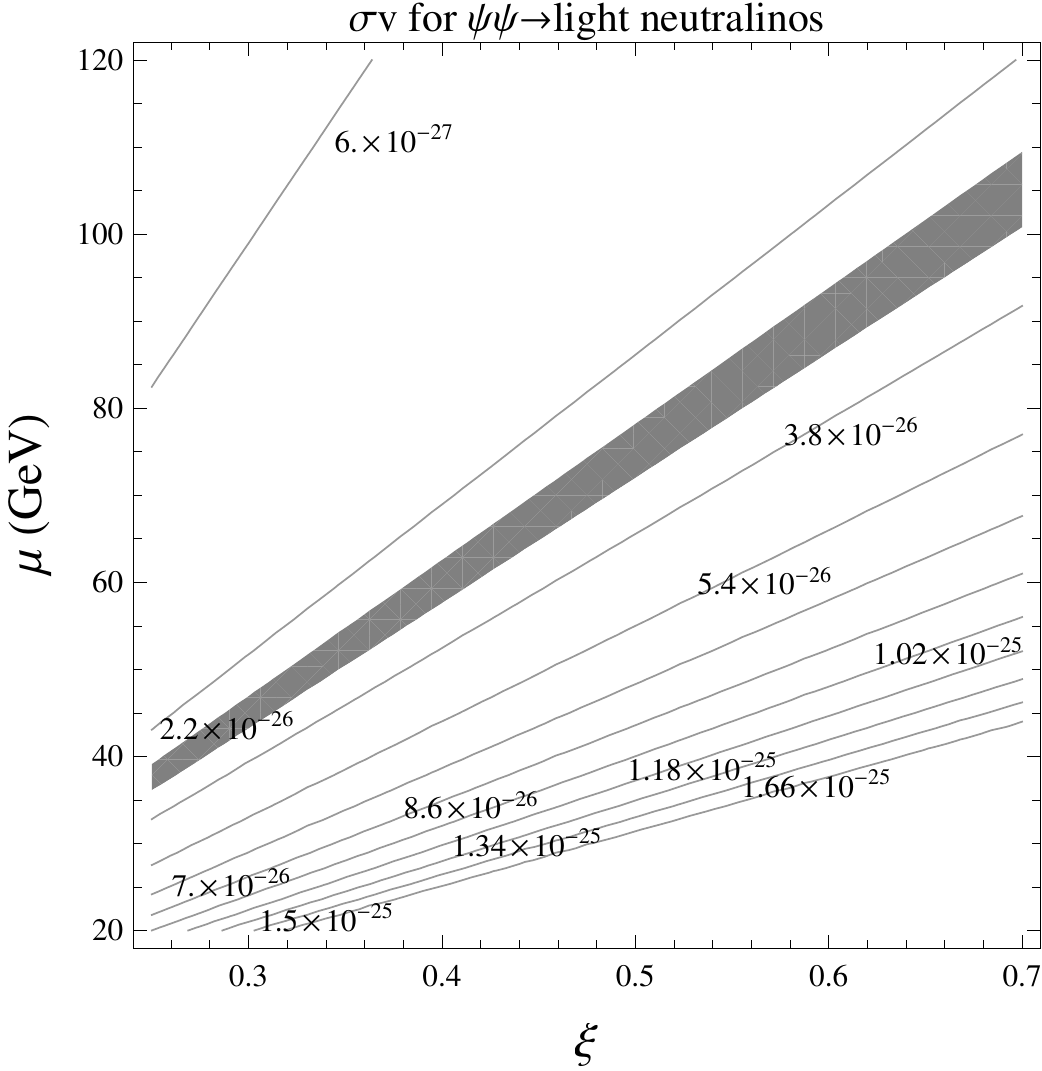}
\caption{$\sigma v$ for the annihilation of $\psi\psi$ to $\tilde{N}_1\tilde{N}_1$ in units of $\mathrm{cm}^3/\mathrm{s}$ with $\lambda=0.1.$ The gray region shows the cross-section that generates the observed relic abundance.}
\label{fig:annsferm}
\end{figure}

\begin{figure}[htpb]
\includegraphics[width=8cm]{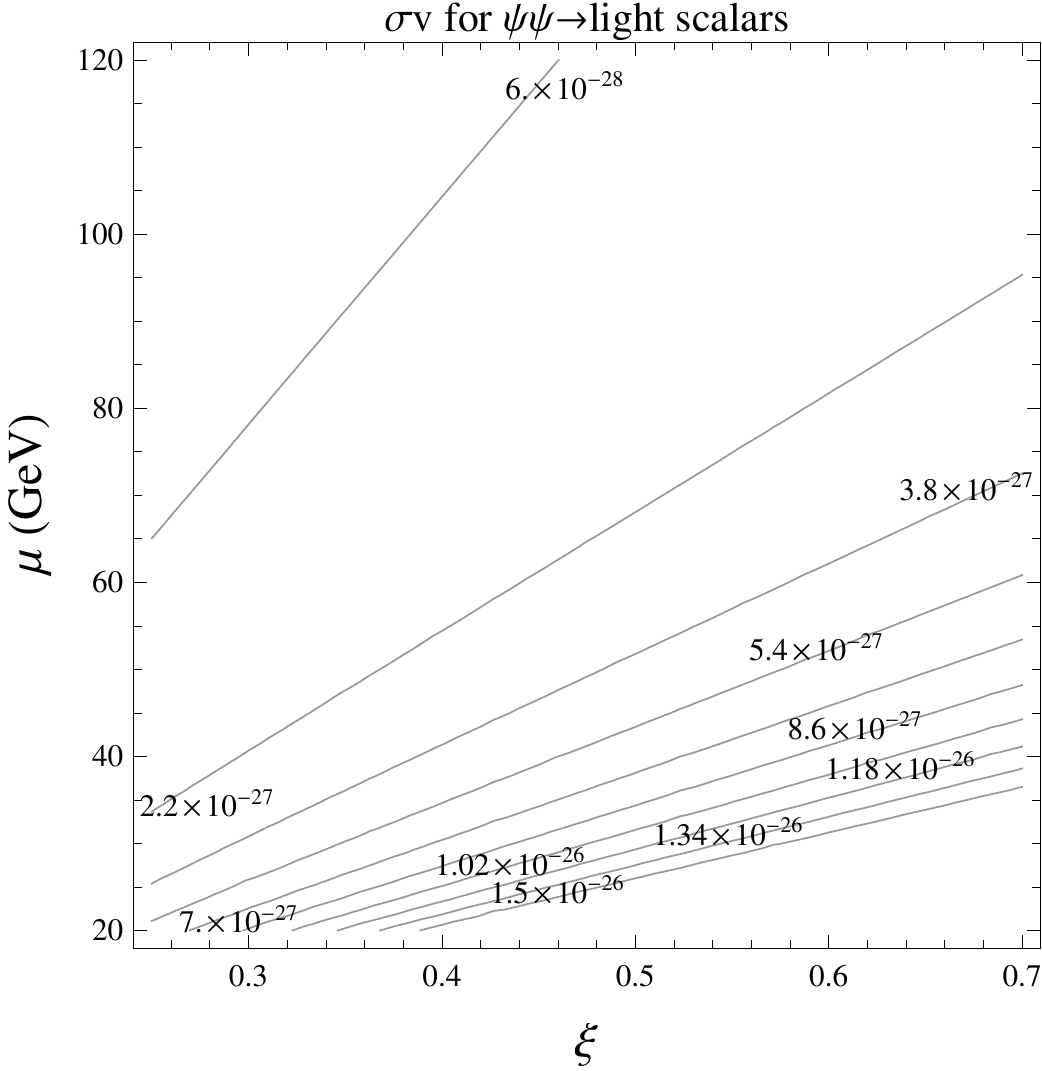}
\caption{$\sigma v$ for the annihilation of $\psi\psi$ to $h_1a_1$ in units of $\mathrm{cm}^3/\mathrm{s}$ with $\lambda=0.1.$}
\label{fig:annscal}
\end{figure}

We implemented this model and evaluated cross-sections within CalcHEP~\cite{Pukhov:2004ca} and found the annihilations are dominated by the process $\psi\psi\rightarrow \tilde{N}_1\tilde{N}_1$. This non-relativistic cross-section, as a function of $\mu$ and $\psi$, is shown in Figure \ref{fig:annsferm}.
The process $\psi\psi\rightarrow h_1a_1$ provides a significant subdominant contribution, approximately 10\% of the previous process. A plot of the cross-section for the same region of parameter space is given in Figure \ref{fig:annscal}.

Higgs production s-channel processes to $a_1h_2$ and $a_1h_3$ are typically suppressed by another factor of $10^2$ from $h_1a_1$ production, and so can be safely ignored for the parameter space of interest.

Finally, we note that in this model the secluded sector may decouple from the other fields of the NMSSM and undergo a separate evolution in the early universe. If there are additional heavy degrees of freedom in the secluded sector that come out of equilibrium during this period, the temperature of the secluded sector will diverge from the temperature of the photons. This shift in temperature may alter the relationship of cross-section and relic density through an effect like the one described in \cite{Nelson:2008hj}. Here $\psi$ and $a_1$ would, respectively, play the role of $X$ and $Y$ from that paper. The $Z_i$ is provided by $\phi_a$ and $\phi_s$, and the temperature difference could be enhanced by including additional heavy unstable singlets.

\subsection{Galactic Annihilations}

In the galaxy, the effective cross-section for dark matter annihilation is increased due to the Sommerfeld effect. In our model, this effect arises predominantly from the exchange of the pseudoscalar $a_1$, as described in \cite{Bedaque:2009ri}.

As described in the preceding section, dark matter annihilation will produce $\tilde{S}$ particles that quickly decay to $a_1$, or will produce $a_1$ and $h_1$ directly. The zero-temperature cross-sections, relevant in the galaxy, for dark matter annihilation to these particles have fairly simple approximate analytic expressions,
\beq
\sigma v_{\psi\psi\rightarrow \tilde{N}_1\tilde{N_1}}&=&\frac{\xi^2\lambda^2}{128\pi\mu^2}+\mathcal{O}(\lambda^4), \\
\sigma v_{\psi\psi\rightarrow h_1a_1}&=&\frac{\xi^2\lambda^2}{2048\pi\mu^2}+\mathcal{O}(\lambda^4).
\eeq
When making numerical calculations, we do not use these approximate expressions, but instead keep all orders of $\lambda$ and $T$.

The $h_1$ and $a_1$ produced in these annihilations have small mixings with the MSSM Higgs that will cause them to decay to the heaviest allowed states. The mass of $a_1$, given (\ref{eq:ma2}), can be kept sub-GeV for sufficiently small $\kappa$, and loop corrections to the mass can be kept under control with a moderately small $\lambda$. This allows kinematic enforcement for the production of light leptons during annihilation, as in the mechanism of \cite{ArkaniHamed:2008qn,Nelson:2008hj} and others. This can give rise to a flux of positrons that is measurable about the modeled astrophysical background, as was observed in the PAMELA experiment~\cite{Adriani:2008zr}.

The $h_1$ typically has a mass above the threshold for baryon production, and so will contribute to the anti-proton flux. However, the cross-section of this subdominant annihilation to baryons is sufficiently small that the excess of anti-protons will lie well within the current experimental uncertainty.

We estimate the positron spectrum within the energy range measured by PAMELA by using approximate solutions to the diffusion equation,
\beq
\dot{f}-K(E)\cdot \bigtriangledown^2f-\frac{\partial}{\partial E}\left(b(E)f\right)=Q,
\eeq
where $f$ is the positron distribution, $K$ is a diffusion constant, $b$ is an energy loss coefficient, and $Q$ is the positron source term. For the case of cylindrical symmetry, an analytical solution for the positron flux observed at Earth is given by
\beq
\Phi(E)=B\frac{v_{e^+}}{8\pi b(E)}\left( \frac{\rho_\odot}{M} \right)^2 \int_E^M dE^\prime \sigma v \frac{dN_{e^+}}{dE_{e^+}} I \left( \lambda_D(E,E^\prime) \right),
\eeq
where K is the diffusion coefficient, $\rho_\odot$ is the dark matter density at our solar system, $I$ is the halo profile, and $\lambda_D$ is the positron diffusion length.
An overview of the approximate solution to this equation in the presence of cylindrical symmetry is given in \cite{Cirelli:2008id}, and we use the ``medium" NFW dark matter profile from that paper.


The electrons and positrons are created at the end of a chain decay. In the dominant process, the $\psi$ annihilation produces two $\tilde{N}_1$ particles, each of which decays to a gravitino and an $a_1$. Depending on its mass, the $a_1$ decays to muons or electrons. We numerically calculate the resulting spectrum, $dN_{e^+}/dE_{e^+}$ by selecting random directions for each step of the decay, isotropic in the decaying particle's rest frame, and then boost the electron 4-momentum to the galactic frame. We perform $10^5$ trails, and fit the resulting spectrum to to the three-parameter function $c_1 E_{e^+}^{c_2}+c_3$, which yields excellent agreement. We perform this procedure for each combination of $\mu$, $\lambda$ and $\xi$ that are used below.

\begin{figure}[htpb]
\includegraphics[width=10cm]{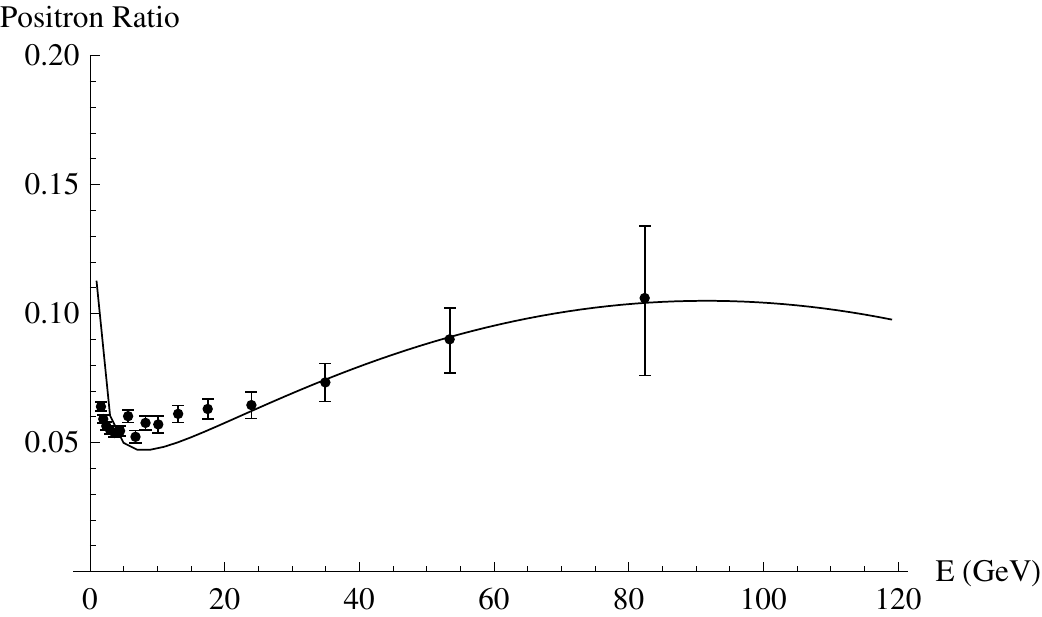}
\caption{The positron spectrum observed near Earth, from the reanalyzed PAMELA data given in \cite{Adriani:2010ib}, shown as a ratio of positrons to the sum of positrons and electrons. The points are the spectrum observed by PAMELA, and the line is the spectrum generated by the present model for the particular choice $\mu=70\GeV$, $\xi=0.3$ and $\lambda=0.1$ (which yields $m_\psi=210\GeV$), and a boost factor of 42.}
\label{fig:pam1}
\end{figure}

\begin{figure}[htpb]
\includegraphics[width=10cm]{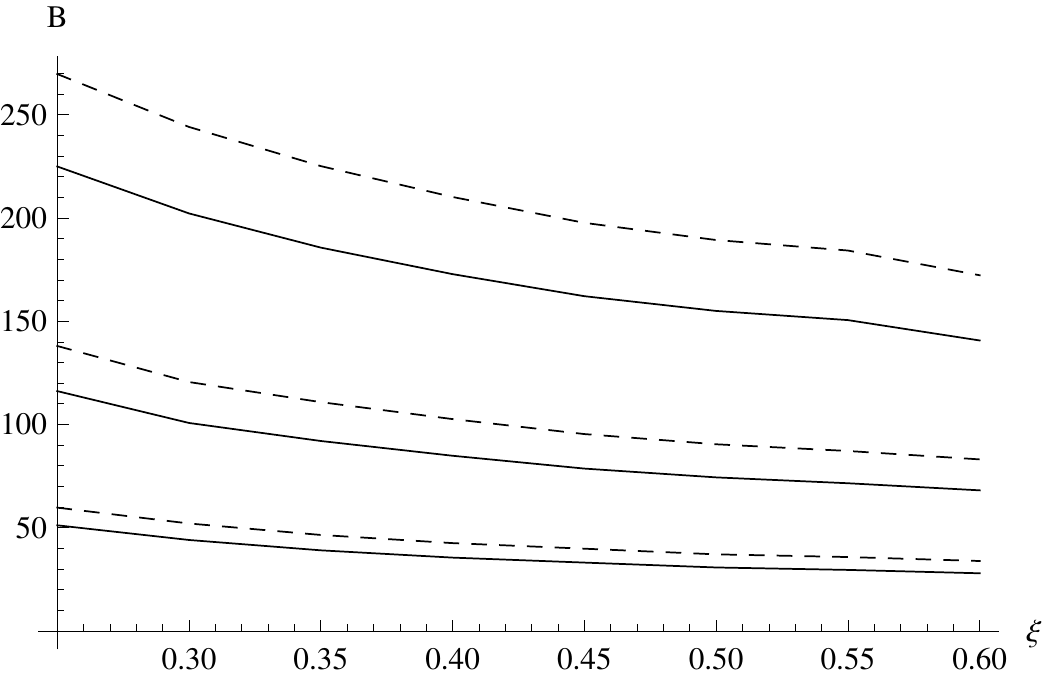}
\caption{The boost factor required, as a function of $\xi$, to fit the model to the PAMELA spectrum for $\lambda=0.1$. The solid curves result from a numerical fit to the newer PAMELA \cite{Adriani:2010ib}, and the dashed curves result from a numerical fit to the original PAMELA spectrum \cite{Adriani:2008zr}. Each neighboring pair of solid and dashed curves is for a different value of $\mu$. From bottom to top, these are $\mu=70,90,110\GeV$. }
\label{fig:boost1}
\end{figure}

For each point in parameter space, we fit the positron spectrum generated by dark matter annihilations to the PAMELA spectrum with an overall scaling ``boost factor." The observed spectrum, originally reported in \cite{Adriani:2008zr}, has recently been updated using a new statistical method \cite{Adriani:2010ib}. The method has resulted in a spectrum that is slightly softer, which suggests a larger dark matter mass and reduced boost factors. We fit our model to both versions of the data.
A typical spectrum that results from the numerical fitting procedure is shown in Figure \ref{fig:pam1}.
The boost factors required may be a combination of enhancement in the local annihilation due to structure within the dark matter halo, and the pseudoscalar Sommerfeld effect.  A plot of the required boost factors across a portion of parameter space is given in Figure \ref{fig:boost1}. The large hierarchy between pseudoscalar mass and dark matter mass, $m_{\tilde{N}_1}/m_{a_1}\sim\mathcal{O}(500)$ easily provides the boost factors required for $\mu\lesssim 100\GeV$.

\section{Discussion and conclusions}
\label{sec:concl}

Our exploration of the dark matter sector in singlet, $\hs$, 
extensions of the MSSM 
shows that explaining the anomalous positron fluxes observed by PAMELA 
with neutralino annihilations is not possible without a carefully tuned choice
of parameters, in regions where the original particle physics motivation
for these extensions is lost.

In the NMSSM, the simplest such extension that addresses the little-fine
tuning and the $\mu$ problems, it is hard to reconcile a neutralino at
the electroweak scale, $m_{\tilde{N}} \gtrsim 100$ GeV, with the light
(pseudo-)scalars that are required to avoid large anti-proton signals,
and to enhance the present day annihilation rate. Indeed, this can only
be achieved by adding non-scale invariant terms to the superpotential. 
These terms could emerge from an unspecified higher energy
theory, but then the scale of the $\mu$ term cannot be explained in the
framework of the NMSSM.

We have showed how adding a dark matter singlet superfield, $\hpsi$, to the
NMSSM in the PQ limit, makes it 
possible to obtain a large positron flux from DM annihilations, without
an unnatural choice of parameters. 
The DM, $\psi$, annihilates primarily into a pair of NMSSM singlino-type
neutralinos. These in turn produce pairs CP-odd higgs, which can be very 
light in the PQ-limit and would mostly generate light leptons upon decay. 
Anti-protons in this scenario are only generated by an annihilation to light scalars that
include the $h_1$, not in the annihilation to neutralinos. Hence the lack
of antiprotons in the spectrum is naturally generated.

A similar scenario had been considered
in~\cite{Nomura:2008ru}, but our setup differs in a few important aspects.   
Our DM is a Majorana fermion, whose stability is ensured by a $\mathbb{Z}_2$
symmetry, that annihilates primarily into a pair of NMSSM singlinos, 
whereas vector-like DM annihilation can be dominantly to 
$h_1+a_1$~\cite{Nomura:2008ru}. Positrons are generated by the
decay of the $a_1$, which appears further down in the reaction chain
in our setup, and resulting in a softer spectrum. The updated positron
spectrum is softer at high energies~\cite{Adriani:2010ib} than
suggested by initial estimates, and we have shown how our model
can fit the data with {\it natural} values of the parameters in 
the DM sector, $\xi \lesssim 1$, and in the electroweak sector, 
$\lambda \lesssim 1$ and $\mu \sim 100$ GeV. The excess in the positron
fraction could extend to higher energies, and there have been conflicting
measurements of a large bump by ATIC~\cite{:2008zzr}, which is not
seen by Fermi~\cite{Abdo:2009zk}. Systematic uncertainties make these
measurements challenging, and more experimental work needs to be
done to clarify the spectrum above the PAMELA cut-off~\cite{Israel:2009}.
In view of these facts, we have not attempted to fit any of the higher
energy experiments in our model. Let us briefly mention, that we can
allow masses for the DM particle that are large enough to create
positron fluxes at much larger energies. This is not the case in other
scenarios, where going beyond the top quark mass is 
problematic~\cite{Bai:2009ka}. We also predict a softer spectrum than
in models where the positrons appear earlier in the annihilation 
chain~\cite{Bai:2009ka,Hooper:2009gm,Nomura:2008ru,ArkaniHamed:2008qn}.

The annihilation rate in the galactic halo should be larger than at the
time of decoupling, and previous proposals have frequently invoked a 
light scalar particle to achieve this 
boost~\cite{Hooper:2009gm,Nomura:2008ru,ArkaniHamed:2008qn,Pospelov:2007mp}.
However, we have explicitly shown in a particular well-motivated
setup that CP-even scalars are not expected to be light enough
to generate the required Sommerfeld enhancement. Unlike 
previous studies, we do not have any
in the spectrum, but we have pointed out that the light CP-odd higgs
can also be used for this purpose. Hence, the relic DM is thermally produced
in this scenario.
While the theory may appear overly general, observations and naturalness
considerations restrict the allowed couplings to a narrow range. 
The coupling $\xi$ of the DM particles to the singlet field S determines
both the annihilation cross-section and the mass of the DM particles. 
It is remarkable that a value $\xi \sim 0.3$ produces the required
${\cal O}(100 \GeV)$ mass, while it also yields the observed relic density.
Hence, insisting on a DM particle that is motivated independently by
particle physics naturalness considerations we have delineated
a viable scenario that can account for the PAMELA observations without
requiring a fine-tuning of the parameters, and addressing at the same time
some of the shortcomings of the MSSM.

\begin{acknowledgments}
The authors are grateful to Yasunori Nomura for useful conversations. This
work was supported in part by the U.S. DOE under Contract
No. DE-FG02-91ER40628 and the NSF under Grant No. PHY-0855580. 
\end{acknowledgments}

\bibliography{dmnmssm}

\begin{thebibliography}{64}
\expandafter\ifx\csname natexlab\endcsname\relax\def\natexlab#1{#1}\fi
\expandafter\ifx\csname bibnamefont\endcsname\relax
  \def\bibnamefont#1{#1}\fi
\expandafter\ifx\csname bibfnamefont\endcsname\relax
  \def\bibfnamefont#1{#1}\fi
\expandafter\ifx\csname citenamefont\endcsname\relax
  \def\citenamefont#1{#1}\fi
\expandafter\ifx\csname url\endcsname\relax
  \def\url#1{\texttt{#1}}\fi
\expandafter\ifx\csname urlprefix\endcsname\relax\def\urlprefix{URL }\fi
\providecommand{\bibinfo}[2]{#2}
\providecommand{\eprint}[2][]{\url{#2}}

\bibitem[{\citenamefont{Komatsu et~al.}(2009)}]{Komatsu:2008hk}
\bibinfo{author}{\bibfnamefont{E.}~\bibnamefont{Komatsu}} \bibnamefont{et~al.}
  (\bibinfo{collaboration}{WMAP}), \bibinfo{journal}{Astrophys. J. Suppl.}
  \textbf{\bibinfo{volume}{180}}, \bibinfo{pages}{330} (\bibinfo{year}{2009}),
  \eprint{0803.0547}.

\bibitem[{\citenamefont{Jungman et~al.}(1996)\citenamefont{Jungman,
  Kamionkowski, and Griest}}]{Jungman:1995df}
\bibinfo{author}{\bibfnamefont{G.}~\bibnamefont{Jungman}},
  \bibinfo{author}{\bibfnamefont{M.}~\bibnamefont{Kamionkowski}},
  \bibnamefont{and} \bibinfo{author}{\bibfnamefont{K.}~\bibnamefont{Griest}},
  \bibinfo{journal}{Phys. Rept.} \textbf{\bibinfo{volume}{267}},
  \bibinfo{pages}{195} (\bibinfo{year}{1996}), \eprint{hep-ph/9506380}.

\bibitem[{\citenamefont{Bertone et~al.}(2005)\citenamefont{Bertone, Hooper, and
  Silk}}]{Bertone:2004pz}
\bibinfo{author}{\bibfnamefont{G.}~\bibnamefont{Bertone}},
  \bibinfo{author}{\bibfnamefont{D.}~\bibnamefont{Hooper}}, \bibnamefont{and}
  \bibinfo{author}{\bibfnamefont{J.}~\bibnamefont{Silk}},
  \bibinfo{journal}{Phys. Rept.} \textbf{\bibinfo{volume}{405}},
  \bibinfo{pages}{279} (\bibinfo{year}{2005}), \eprint{hep-ph/0404175}.

\bibitem[{\citenamefont{Adriani et~al.}(2009{\natexlab{a}})}]{Adriani:2008zr}
\bibinfo{author}{\bibfnamefont{O.}~\bibnamefont{Adriani}} \bibnamefont{et~al.}
  (\bibinfo{collaboration}{PAMELA}), \bibinfo{journal}{Nature}
  \textbf{\bibinfo{volume}{458}}, \bibinfo{pages}{607}
  (\bibinfo{year}{2009}{\natexlab{a}}), \eprint{0810.4995}.

\bibitem[{\citenamefont{Adriani et~al.}(2010)}]{Adriani:2010ib}
\bibinfo{author}{\bibfnamefont{O.}~\bibnamefont{Adriani}} \bibnamefont{et~al.}
  (\bibinfo{year}{2010}), \eprint{1001.3522}.

\bibitem[{\citenamefont{Cirelli and Strumia}(2008)}]{Cirelli:2008jk}
\bibinfo{author}{\bibfnamefont{M.}~\bibnamefont{Cirelli}} \bibnamefont{and}
  \bibinfo{author}{\bibfnamefont{A.}~\bibnamefont{Strumia}}
  (\bibinfo{year}{2008}), \eprint{0808.3867}.

\bibitem[{\citenamefont{Barger et~al.}(2009)\citenamefont{Barger, Keung,
  Marfatia, and Shaughnessy}}]{Barger:2008su}
\bibinfo{author}{\bibfnamefont{V.}~\bibnamefont{Barger}},
  \bibinfo{author}{\bibfnamefont{W.~Y.} \bibnamefont{Keung}},
  \bibinfo{author}{\bibfnamefont{D.}~\bibnamefont{Marfatia}}, \bibnamefont{and}
  \bibinfo{author}{\bibfnamefont{G.}~\bibnamefont{Shaughnessy}},
  \bibinfo{journal}{Phys. Lett.} \textbf{\bibinfo{volume}{B672}},
  \bibinfo{pages}{141} (\bibinfo{year}{2009}), \eprint{0809.0162}.

\bibitem[{\citenamefont{Cholis et~al.}(2009)\citenamefont{Cholis, Goodenough,
  Hooper, Simet, and Weiner}}]{Cholis:2008hb}
\bibinfo{author}{\bibfnamefont{I.}~\bibnamefont{Cholis}},
  \bibinfo{author}{\bibfnamefont{L.}~\bibnamefont{Goodenough}},
  \bibinfo{author}{\bibfnamefont{D.}~\bibnamefont{Hooper}},
  \bibinfo{author}{\bibfnamefont{M.}~\bibnamefont{Simet}}, \bibnamefont{and}
  \bibinfo{author}{\bibfnamefont{N.}~\bibnamefont{Weiner}},
  \bibinfo{journal}{Phys. Rev.} \textbf{\bibinfo{volume}{D80}},
  \bibinfo{pages}{123511} (\bibinfo{year}{2009}), \eprint{0809.1683}.

\bibitem[{\citenamefont{Arkani-Hamed et~al.}(2009)\citenamefont{Arkani-Hamed,
  Finkbeiner, Slatyer, and Weiner}}]{ArkaniHamed:2008qn}
\bibinfo{author}{\bibfnamefont{N.}~\bibnamefont{Arkani-Hamed}},
  \bibinfo{author}{\bibfnamefont{D.~P.} \bibnamefont{Finkbeiner}},
  \bibinfo{author}{\bibfnamefont{T.~R.} \bibnamefont{Slatyer}},
  \bibnamefont{and} \bibinfo{author}{\bibfnamefont{N.}~\bibnamefont{Weiner}},
  \bibinfo{journal}{Phys. Rev.} \textbf{\bibinfo{volume}{D79}},
  \bibinfo{pages}{015014} (\bibinfo{year}{2009}), \eprint{0810.0713}.

\bibitem[{\citenamefont{Pospelov and Ritz}(2009)}]{Pospelov:2008jd}
\bibinfo{author}{\bibfnamefont{M.}~\bibnamefont{Pospelov}} \bibnamefont{and}
  \bibinfo{author}{\bibfnamefont{A.}~\bibnamefont{Ritz}},
  \bibinfo{journal}{Phys. Lett.} \textbf{\bibinfo{volume}{B671}},
  \bibinfo{pages}{391} (\bibinfo{year}{2009}), \eprint{0810.1502}.

\bibitem[{\citenamefont{Nelson and Spitzer}(2008)}]{Nelson:2008hj}
\bibinfo{author}{\bibfnamefont{A.~E.} \bibnamefont{Nelson}} \bibnamefont{and}
  \bibinfo{author}{\bibfnamefont{C.}~\bibnamefont{Spitzer}}
  (\bibinfo{year}{2008}), \eprint{0810.5167}.

\bibitem[{\citenamefont{Nomura and Thaler}(2009)}]{Nomura:2008ru}
\bibinfo{author}{\bibfnamefont{Y.}~\bibnamefont{Nomura}} \bibnamefont{and}
  \bibinfo{author}{\bibfnamefont{J.}~\bibnamefont{Thaler}},
  \bibinfo{journal}{Phys. Rev.} \textbf{\bibinfo{volume}{D79}},
  \bibinfo{pages}{075008} (\bibinfo{year}{2009}), \eprint{0810.5397}.

\bibitem[{\citenamefont{Harnik and Kribs}(2009)}]{Harnik:2008uu}
\bibinfo{author}{\bibfnamefont{R.}~\bibnamefont{Harnik}} \bibnamefont{and}
  \bibinfo{author}{\bibfnamefont{G.~D.} \bibnamefont{Kribs}},
  \bibinfo{journal}{Phys. Rev.} \textbf{\bibinfo{volume}{D79}},
  \bibinfo{pages}{095007} (\bibinfo{year}{2009}), \eprint{0810.5557}.

\bibitem[{\citenamefont{Fox and Poppitz}(2009)}]{Fox:2008kb}
\bibinfo{author}{\bibfnamefont{P.~J.} \bibnamefont{Fox}} \bibnamefont{and}
  \bibinfo{author}{\bibfnamefont{E.}~\bibnamefont{Poppitz}},
  \bibinfo{journal}{Phys. Rev.} \textbf{\bibinfo{volume}{D79}},
  \bibinfo{pages}{083528} (\bibinfo{year}{2009}), \eprint{0811.0399}.

\bibitem[{\citenamefont{Hooper and Zurek}(2009)}]{Hooper:2009fj}
\bibinfo{author}{\bibfnamefont{D.}~\bibnamefont{Hooper}} \bibnamefont{and}
  \bibinfo{author}{\bibfnamefont{K.~M.} \bibnamefont{Zurek}},
  \bibinfo{journal}{Phys. Rev.} \textbf{\bibinfo{volume}{D79}},
  \bibinfo{pages}{103529} (\bibinfo{year}{2009}), \eprint{0902.0593}.

\bibitem[{\citenamefont{Bai et~al.}(2009)\citenamefont{Bai, Carena, and
  Lykken}}]{Bai:2009ka}
\bibinfo{author}{\bibfnamefont{Y.}~\bibnamefont{Bai}},
  \bibinfo{author}{\bibfnamefont{M.}~\bibnamefont{Carena}}, \bibnamefont{and}
  \bibinfo{author}{\bibfnamefont{J.}~\bibnamefont{Lykken}},
  \bibinfo{journal}{Phys. Rev.} \textbf{\bibinfo{volume}{D80}},
  \bibinfo{pages}{055004} (\bibinfo{year}{2009}), \eprint{0905.2964}.

\bibitem[{\citenamefont{Hooper and Tait}(2009)}]{Hooper:2009gm}
\bibinfo{author}{\bibfnamefont{D.}~\bibnamefont{Hooper}} \bibnamefont{and}
  \bibinfo{author}{\bibfnamefont{T.~M.~P.} \bibnamefont{Tait}}
  (\bibinfo{year}{2009}), \eprint{0906.0362}.

\bibitem[{\citenamefont{Wang et~al.}(2009)\citenamefont{Wang, Xiong, Yang, and
  Yu}}]{Wang:2009rj}
\bibinfo{author}{\bibfnamefont{W.}~\bibnamefont{Wang}},
  \bibinfo{author}{\bibfnamefont{Z.}~\bibnamefont{Xiong}},
  \bibinfo{author}{\bibfnamefont{J.~M.} \bibnamefont{Yang}}, \bibnamefont{and}
  \bibinfo{author}{\bibfnamefont{L.-X.} \bibnamefont{Yu}}
  (\bibinfo{year}{2009}), \eprint{0908.0486}.

\bibitem[{\citenamefont{Park and Shu}(2009)}]{Park:2009cs}
\bibinfo{author}{\bibfnamefont{S.~C.} \bibnamefont{Park}} \bibnamefont{and}
  \bibinfo{author}{\bibfnamefont{J.}~\bibnamefont{Shu}},
  \bibinfo{journal}{Phys. Rev.} \textbf{\bibinfo{volume}{D79}},
  \bibinfo{pages}{091702} (\bibinfo{year}{2009}), \eprint{0901.0720}.

\bibitem[{\citenamefont{Yin et~al.}(2009)}]{Yin:2008bs}
\bibinfo{author}{\bibfnamefont{P.-f.} \bibnamefont{Yin}} \bibnamefont{et~al.},
  \bibinfo{journal}{Phys. Rev.} \textbf{\bibinfo{volume}{D79}},
  \bibinfo{pages}{023512} (\bibinfo{year}{2009}), \eprint{0811.0176}.

\bibitem[{\citenamefont{Nardi et~al.}(2009)\citenamefont{Nardi, Sannino, and
  Strumia}}]{Nardi:2008ix}
\bibinfo{author}{\bibfnamefont{E.}~\bibnamefont{Nardi}},
  \bibinfo{author}{\bibfnamefont{F.}~\bibnamefont{Sannino}}, \bibnamefont{and}
  \bibinfo{author}{\bibfnamefont{A.}~\bibnamefont{Strumia}},
  \bibinfo{journal}{JCAP} \textbf{\bibinfo{volume}{0901}}, \bibinfo{pages}{043}
  (\bibinfo{year}{2009}), \eprint{0811.4153}.

\bibitem[{\citenamefont{Ibarra and Tran}(2009)}]{Ibarra:2008jk}
\bibinfo{author}{\bibfnamefont{A.}~\bibnamefont{Ibarra}} \bibnamefont{and}
  \bibinfo{author}{\bibfnamefont{D.}~\bibnamefont{Tran}},
  \bibinfo{journal}{JCAP} \textbf{\bibinfo{volume}{0902}}, \bibinfo{pages}{021}
  (\bibinfo{year}{2009}), \eprint{0811.1555}.

\bibitem[{\citenamefont{Ishiwata et~al.}(2009)\citenamefont{Ishiwata,
  Matsumoto, and Moroi}}]{Ishiwata:2009vx}
\bibinfo{author}{\bibfnamefont{K.}~\bibnamefont{Ishiwata}},
  \bibinfo{author}{\bibfnamefont{S.}~\bibnamefont{Matsumoto}},
  \bibnamefont{and} \bibinfo{author}{\bibfnamefont{T.}~\bibnamefont{Moroi}},
  \bibinfo{journal}{JHEP} \textbf{\bibinfo{volume}{05}}, \bibinfo{pages}{110}
  (\bibinfo{year}{2009}), \eprint{0903.0242}.

\bibitem[{\citenamefont{Chen et~al.}(2009)\citenamefont{Chen, Mohapatra,
  Nussinov, and Zhang}}]{Chen:2009ew}
\bibinfo{author}{\bibfnamefont{S.-L.} \bibnamefont{Chen}},
  \bibinfo{author}{\bibfnamefont{R.~N.} \bibnamefont{Mohapatra}},
  \bibinfo{author}{\bibfnamefont{S.}~\bibnamefont{Nussinov}}, \bibnamefont{and}
  \bibinfo{author}{\bibfnamefont{Y.}~\bibnamefont{Zhang}},
  \bibinfo{journal}{Phys. Lett.} \textbf{\bibinfo{volume}{B677}},
  \bibinfo{pages}{311} (\bibinfo{year}{2009}), \eprint{0903.2562}.

\bibitem[{\citenamefont{Arvanitaki et~al.}(2009)}]{Arvanitaki:2009yb}
\bibinfo{author}{\bibfnamefont{A.}~\bibnamefont{Arvanitaki}}
  \bibnamefont{et~al.}, \bibinfo{journal}{Phys. Rev.}
  \textbf{\bibinfo{volume}{D80}}, \bibinfo{pages}{055011}
  (\bibinfo{year}{2009}), \eprint{0904.2789}.

\bibitem[{\citenamefont{Hooper et~al.}(2009{\natexlab{a}})\citenamefont{Hooper,
  Blasi, and Serpico}}]{Hooper:2008kg}
\bibinfo{author}{\bibfnamefont{D.}~\bibnamefont{Hooper}},
  \bibinfo{author}{\bibfnamefont{P.}~\bibnamefont{Blasi}}, \bibnamefont{and}
  \bibinfo{author}{\bibfnamefont{P.~D.} \bibnamefont{Serpico}},
  \bibinfo{journal}{JCAP} \textbf{\bibinfo{volume}{0901}}, \bibinfo{pages}{025}
  (\bibinfo{year}{2009}{\natexlab{a}}), \eprint{0810.1527}.

\bibitem[{\citenamefont{Yuksel et~al.}(2009)\citenamefont{Yuksel, Kistler, and
  Stanev}}]{Yuksel:2008rf}
\bibinfo{author}{\bibfnamefont{H.}~\bibnamefont{Yuksel}},
  \bibinfo{author}{\bibfnamefont{M.~D.} \bibnamefont{Kistler}},
  \bibnamefont{and} \bibinfo{author}{\bibfnamefont{T.}~\bibnamefont{Stanev}},
  \bibinfo{journal}{Phys. Rev. Lett.} \textbf{\bibinfo{volume}{103}},
  \bibinfo{pages}{051101} (\bibinfo{year}{2009}), \eprint{0810.2784}.

\bibitem[{\citenamefont{Profumo}(2008)}]{Profumo:2008ms}
\bibinfo{author}{\bibfnamefont{S.}~\bibnamefont{Profumo}}
  (\bibinfo{year}{2008}), \eprint{0812.4457}.

\bibitem[{\citenamefont{Shaviv et~al.}(2009)\citenamefont{Shaviv, Nakar, and
  Piran}}]{Shaviv:2009bu}
\bibinfo{author}{\bibfnamefont{N.~J.} \bibnamefont{Shaviv}},
  \bibinfo{author}{\bibfnamefont{E.}~\bibnamefont{Nakar}}, \bibnamefont{and}
  \bibinfo{author}{\bibfnamefont{T.}~\bibnamefont{Piran}},
  \bibinfo{journal}{Phys. Rev. Lett.} \textbf{\bibinfo{volume}{103}},
  \bibinfo{pages}{111302} (\bibinfo{year}{2009}), \eprint{0902.0376}.

\bibitem[{\citenamefont{Blasi and Serpico}(2009)}]{Blasi:2009bd}
\bibinfo{author}{\bibfnamefont{P.}~\bibnamefont{Blasi}} \bibnamefont{and}
  \bibinfo{author}{\bibfnamefont{P.~D.} \bibnamefont{Serpico}},
  \bibinfo{journal}{Phys. Rev. Lett.} \textbf{\bibinfo{volume}{103}},
  \bibinfo{pages}{081103} (\bibinfo{year}{2009}), \eprint{0904.0871}.

\bibitem[{\citenamefont{Blasi}(2009)}]{Blasi:2009hv}
\bibinfo{author}{\bibfnamefont{P.}~\bibnamefont{Blasi}},
  \bibinfo{journal}{Phys. Rev. Lett.} \textbf{\bibinfo{volume}{103}},
  \bibinfo{pages}{051104} (\bibinfo{year}{2009}), \eprint{0903.2794}.

\bibitem[{\citenamefont{Cowsik and Burch}(2009)}]{Cowsik:2009ga}
\bibinfo{author}{\bibfnamefont{R.}~\bibnamefont{Cowsik}} \bibnamefont{and}
  \bibinfo{author}{\bibfnamefont{B.}~\bibnamefont{Burch}}
  (\bibinfo{year}{2009}), \eprint{0905.2136}.

\bibitem[{\citenamefont{Mertsch and Sarkar}(2009)}]{Mertsch:2009ph}
\bibinfo{author}{\bibfnamefont{P.}~\bibnamefont{Mertsch}} \bibnamefont{and}
  \bibinfo{author}{\bibfnamefont{S.}~\bibnamefont{Sarkar}},
  \bibinfo{journal}{Phys. Rev. Lett.} \textbf{\bibinfo{volume}{103}},
  \bibinfo{pages}{081104} (\bibinfo{year}{2009}), \eprint{0905.3152}.

\bibitem[{\citenamefont{Ahlers et~al.}(2009)\citenamefont{Ahlers, Mertsch, and
  Sarkar}}]{Ahlers:2009ae}
\bibinfo{author}{\bibfnamefont{M.}~\bibnamefont{Ahlers}},
  \bibinfo{author}{\bibfnamefont{P.}~\bibnamefont{Mertsch}}, \bibnamefont{and}
  \bibinfo{author}{\bibfnamefont{S.}~\bibnamefont{Sarkar}},
  \bibinfo{journal}{Phys. Rev.} \textbf{\bibinfo{volume}{D80}},
  \bibinfo{pages}{123017} (\bibinfo{year}{2009}), \eprint{0909.4060}.

\bibitem[{\citenamefont{Adriani et~al.}(2009{\natexlab{b}})}]{Adriani:2008zq}
\bibinfo{author}{\bibfnamefont{O.}~\bibnamefont{Adriani}} \bibnamefont{et~al.},
  \bibinfo{journal}{Phys. Rev. Lett.} \textbf{\bibinfo{volume}{102}},
  \bibinfo{pages}{051101} (\bibinfo{year}{2009}{\natexlab{b}}),
  \eprint{0810.4994}.

\bibitem[{\citenamefont{Lee and Weinberg}(1977)}]{Lee:1977ua}
\bibinfo{author}{\bibfnamefont{B.~W.} \bibnamefont{Lee}} \bibnamefont{and}
  \bibinfo{author}{\bibfnamefont{S.}~\bibnamefont{Weinberg}},
  \bibinfo{journal}{Phys. Rev. Lett.} \textbf{\bibinfo{volume}{39}},
  \bibinfo{pages}{165} (\bibinfo{year}{1977}).

\bibitem[{\citenamefont{Bergstrom et~al.}(2008)\citenamefont{Bergstrom,
  Bringmann, and Edsjo}}]{Bergstrom:2008gr}
\bibinfo{author}{\bibfnamefont{L.}~\bibnamefont{Bergstrom}},
  \bibinfo{author}{\bibfnamefont{T.}~\bibnamefont{Bringmann}},
  \bibnamefont{and} \bibinfo{author}{\bibfnamefont{J.}~\bibnamefont{Edsjo}},
  \bibinfo{journal}{Phys. Rev.} \textbf{\bibinfo{volume}{D78}},
  \bibinfo{pages}{103520} (\bibinfo{year}{2008}), \eprint{0808.3725}.

\bibitem[{\citenamefont{Grajek et~al.}(2009)\citenamefont{Grajek, Kane, Phalen,
  Pierce, and Watson}}]{Grajek:2008pg}
\bibinfo{author}{\bibfnamefont{P.}~\bibnamefont{Grajek}},
  \bibinfo{author}{\bibfnamefont{G.}~\bibnamefont{Kane}},
  \bibinfo{author}{\bibfnamefont{D.}~\bibnamefont{Phalen}},
  \bibinfo{author}{\bibfnamefont{A.}~\bibnamefont{Pierce}}, \bibnamefont{and}
  \bibinfo{author}{\bibfnamefont{S.}~\bibnamefont{Watson}},
  \bibinfo{journal}{Phys. Rev.} \textbf{\bibinfo{volume}{D79}},
  \bibinfo{pages}{043506} (\bibinfo{year}{2009}), \eprint{0812.4555}.

\bibitem[{\citenamefont{Hooper et~al.}(2009{\natexlab{b}})\citenamefont{Hooper,
  Stebbins, and Zurek}}]{Hooper:2008kv}
\bibinfo{author}{\bibfnamefont{D.}~\bibnamefont{Hooper}},
  \bibinfo{author}{\bibfnamefont{A.}~\bibnamefont{Stebbins}}, \bibnamefont{and}
  \bibinfo{author}{\bibfnamefont{K.~M.} \bibnamefont{Zurek}},
  \bibinfo{journal}{Phys. Rev.} \textbf{\bibinfo{volume}{D79}},
  \bibinfo{pages}{103513} (\bibinfo{year}{2009}{\natexlab{b}}),
  \eprint{0812.3202}.

\bibitem[{\citenamefont{Brun et~al.}(2009)\citenamefont{Brun, Delahaye,
  Diemand, Profumo, and Salati}}]{Brun:2009aj}
\bibinfo{author}{\bibfnamefont{P.}~\bibnamefont{Brun}},
  \bibinfo{author}{\bibfnamefont{T.}~\bibnamefont{Delahaye}},
  \bibinfo{author}{\bibfnamefont{J.}~\bibnamefont{Diemand}},
  \bibinfo{author}{\bibfnamefont{S.}~\bibnamefont{Profumo}}, \bibnamefont{and}
  \bibinfo{author}{\bibfnamefont{P.}~\bibnamefont{Salati}},
  \bibinfo{journal}{Phys. Rev.} \textbf{\bibinfo{volume}{D80}},
  \bibinfo{pages}{035023} (\bibinfo{year}{2009}), \eprint{0904.0812}.

\bibitem[{\citenamefont{Gogoladze et~al.}(2009)\citenamefont{Gogoladze, Khalid,
  Shafi, and Yuksel}}]{Gogoladze:2009kv}
\bibinfo{author}{\bibfnamefont{I.}~\bibnamefont{Gogoladze}},
  \bibinfo{author}{\bibfnamefont{R.}~\bibnamefont{Khalid}},
  \bibinfo{author}{\bibfnamefont{Q.}~\bibnamefont{Shafi}}, \bibnamefont{and}
  \bibinfo{author}{\bibfnamefont{H.}~\bibnamefont{Yuksel}},
  \bibinfo{journal}{Phys. Rev.} \textbf{\bibinfo{volume}{D79}},
  \bibinfo{pages}{055019} (\bibinfo{year}{2009}), \eprint{0901.0923}.

\bibitem[{\citenamefont{Hisano et~al.}(2005)\citenamefont{Hisano, Matsumoto,
  Nojiri, and Saito}}]{Hisano:2004ds}
\bibinfo{author}{\bibfnamefont{J.}~\bibnamefont{Hisano}},
  \bibinfo{author}{\bibfnamefont{S.}~\bibnamefont{Matsumoto}},
  \bibinfo{author}{\bibfnamefont{M.~M.} \bibnamefont{Nojiri}},
  \bibnamefont{and} \bibinfo{author}{\bibfnamefont{O.}~\bibnamefont{Saito}},
  \bibinfo{journal}{Phys. Rev.} \textbf{\bibinfo{volume}{D71}},
  \bibinfo{pages}{063528} (\bibinfo{year}{2005}), \eprint{hep-ph/0412403}.

\bibitem[{\citenamefont{Cirelli et~al.}(2007)\citenamefont{Cirelli, Strumia,
  and Tamburini}}]{Cirelli:2007xd}
\bibinfo{author}{\bibfnamefont{M.}~\bibnamefont{Cirelli}},
  \bibinfo{author}{\bibfnamefont{A.}~\bibnamefont{Strumia}}, \bibnamefont{and}
  \bibinfo{author}{\bibfnamefont{M.}~\bibnamefont{Tamburini}},
  \bibinfo{journal}{Nucl. Phys.} \textbf{\bibinfo{volume}{B787}},
  \bibinfo{pages}{152} (\bibinfo{year}{2007}), \eprint{0706.4071}.

\bibitem[{\citenamefont{Pospelov et~al.}(2008)\citenamefont{Pospelov, Ritz, and
  Voloshin}}]{Pospelov:2007mp}
\bibinfo{author}{\bibfnamefont{M.}~\bibnamefont{Pospelov}},
  \bibinfo{author}{\bibfnamefont{A.}~\bibnamefont{Ritz}}, \bibnamefont{and}
  \bibinfo{author}{\bibfnamefont{M.~B.} \bibnamefont{Voloshin}},
  \bibinfo{journal}{Phys. Lett.} \textbf{\bibinfo{volume}{B662}},
  \bibinfo{pages}{53} (\bibinfo{year}{2008}), \eprint{0711.4866}.

\bibitem[{\citenamefont{Kim and Nilles}(1984)}]{Kim:1983dt}
\bibinfo{author}{\bibfnamefont{J.~E.} \bibnamefont{Kim}} \bibnamefont{and}
  \bibinfo{author}{\bibfnamefont{H.~P.} \bibnamefont{Nilles}},
  \bibinfo{journal}{Phys. Lett.} \textbf{\bibinfo{volume}{B138}},
  \bibinfo{pages}{150} (\bibinfo{year}{1984}).

\bibitem[{\citenamefont{Bastero-Gil
  et~al.}(2000{\natexlab{a}})\citenamefont{Bastero-Gil, Kane, and
  King}}]{BasteroGil:1999gu}
\bibinfo{author}{\bibfnamefont{M.}~\bibnamefont{Bastero-Gil}},
  \bibinfo{author}{\bibfnamefont{G.~L.} \bibnamefont{Kane}}, \bibnamefont{and}
  \bibinfo{author}{\bibfnamefont{S.~F.} \bibnamefont{King}},
  \bibinfo{journal}{Phys. Lett.} \textbf{\bibinfo{volume}{B474}},
  \bibinfo{pages}{103} (\bibinfo{year}{2000}{\natexlab{a}}),
  \eprint{hep-ph/9910506}.

\bibitem[{\citenamefont{Bastero-Gil
  et~al.}(2000{\natexlab{b}})\citenamefont{Bastero-Gil, Hugonie, King, Roy, and
  Vempati}}]{BasteroGil:2000bw}
\bibinfo{author}{\bibfnamefont{M.}~\bibnamefont{Bastero-Gil}},
  \bibinfo{author}{\bibfnamefont{C.}~\bibnamefont{Hugonie}},
  \bibinfo{author}{\bibfnamefont{S.~F.} \bibnamefont{King}},
  \bibinfo{author}{\bibfnamefont{D.~P.} \bibnamefont{Roy}}, \bibnamefont{and}
  \bibinfo{author}{\bibfnamefont{S.}~\bibnamefont{Vempati}},
  \bibinfo{journal}{Phys. Lett.} \textbf{\bibinfo{volume}{B489}},
  \bibinfo{pages}{359} (\bibinfo{year}{2000}{\natexlab{b}}),
  \eprint{hep-ph/0006198}.

\bibitem[{\citenamefont{Dermisek and Gunion}(2005)}]{Dermisek:2005ar}
\bibinfo{author}{\bibfnamefont{R.}~\bibnamefont{Dermisek}} \bibnamefont{and}
  \bibinfo{author}{\bibfnamefont{J.~F.} \bibnamefont{Gunion}},
  \bibinfo{journal}{Phys. Rev. Lett.} \textbf{\bibinfo{volume}{95}},
  \bibinfo{pages}{041801} (\bibinfo{year}{2005}), \eprint{hep-ph/0502105}.

\bibitem[{\citenamefont{Gunion et~al.}(1996)\citenamefont{Gunion, Haber, and
  Moroi}}]{Gunion:1996fb}
\bibinfo{author}{\bibfnamefont{J.~F.} \bibnamefont{Gunion}},
  \bibinfo{author}{\bibfnamefont{H.~E.} \bibnamefont{Haber}}, \bibnamefont{and}
  \bibinfo{author}{\bibfnamefont{T.}~\bibnamefont{Moroi}}
  (\bibinfo{year}{1996}), \eprint{hep-ph/9610337}.

\bibitem[{\citenamefont{Dobrescu et~al.}(2001)\citenamefont{Dobrescu,
  Landsberg, and Matchev}}]{Dobrescu:2000jt}
\bibinfo{author}{\bibfnamefont{B.~A.} \bibnamefont{Dobrescu}},
  \bibinfo{author}{\bibfnamefont{G.~L.} \bibnamefont{Landsberg}},
  \bibnamefont{and} \bibinfo{author}{\bibfnamefont{K.~T.}
  \bibnamefont{Matchev}}, \bibinfo{journal}{Phys. Rev.}
  \textbf{\bibinfo{volume}{D63}}, \bibinfo{pages}{075003}
  (\bibinfo{year}{2001}), \eprint{hep-ph/0005308}.

\bibitem[{\citenamefont{Maniatis}(2009)}]{Maniatis:2009re}
\bibinfo{author}{\bibfnamefont{M.}~\bibnamefont{Maniatis}}
  (\bibinfo{year}{2009}), \eprint{0906.0777}.

\bibitem[{\citenamefont{Ellwanger et~al.}(2009)\citenamefont{Ellwanger,
  Hugonie, and Teixeira}}]{Ellwanger:2009dp}
\bibinfo{author}{\bibfnamefont{U.}~\bibnamefont{Ellwanger}},
  \bibinfo{author}{\bibfnamefont{C.}~\bibnamefont{Hugonie}}, \bibnamefont{and}
  \bibinfo{author}{\bibfnamefont{A.~M.} \bibnamefont{Teixeira}}
  (\bibinfo{year}{2009}), \eprint{0910.1785}.

\bibitem[{\citenamefont{Ellwanger and Hugonie}(2002)}]{Ellwanger:1999ji}
\bibinfo{author}{\bibfnamefont{U.}~\bibnamefont{Ellwanger}} \bibnamefont{and}
  \bibinfo{author}{\bibfnamefont{C.}~\bibnamefont{Hugonie}},
  \bibinfo{journal}{Eur. Phys. J.} \textbf{\bibinfo{volume}{C25}},
  \bibinfo{pages}{297} (\bibinfo{year}{2002}), \eprint{hep-ph/9909260}.

\bibitem[{\citenamefont{Ferrer et~al.}(2006)\citenamefont{Ferrer, Krauss, and
  Profumo}}]{Ferrer:2006hy}
\bibinfo{author}{\bibfnamefont{F.}~\bibnamefont{Ferrer}},
  \bibinfo{author}{\bibfnamefont{L.~M.} \bibnamefont{Krauss}},
  \bibnamefont{and} \bibinfo{author}{\bibfnamefont{S.}~\bibnamefont{Profumo}},
  \bibinfo{journal}{Phys. Rev.} \textbf{\bibinfo{volume}{D74}},
  \bibinfo{pages}{115007} (\bibinfo{year}{2006}), \eprint{hep-ph/0609257}.

\bibitem[{\citenamefont{Ellwanger and Hugonie}(2006)}]{Ellwanger:2005dv}
\bibinfo{author}{\bibfnamefont{U.}~\bibnamefont{Ellwanger}} \bibnamefont{and}
  \bibinfo{author}{\bibfnamefont{C.}~\bibnamefont{Hugonie}},
  \bibinfo{journal}{Comput. Phys. Commun.} \textbf{\bibinfo{volume}{175}},
  \bibinfo{pages}{290} (\bibinfo{year}{2006}), \eprint{hep-ph/0508022}.

\bibitem[{\citenamefont{Belanger et~al.}(2009)\citenamefont{Belanger, Boudjema,
  Pukhov, and Semenov}}]{Belanger:2008sj}
\bibinfo{author}{\bibfnamefont{G.}~\bibnamefont{Belanger}},
  \bibinfo{author}{\bibfnamefont{F.}~\bibnamefont{Boudjema}},
  \bibinfo{author}{\bibfnamefont{A.}~\bibnamefont{Pukhov}}, \bibnamefont{and}
  \bibinfo{author}{\bibfnamefont{A.}~\bibnamefont{Semenov}},
  \bibinfo{journal}{Comput. Phys. Commun.} \textbf{\bibinfo{volume}{180}},
  \bibinfo{pages}{747} (\bibinfo{year}{2009}), \eprint{0803.2360}.

\bibitem[{\citenamefont{Ferrer and Grifols}(1998)}]{Ferrer:1998ue}
\bibinfo{author}{\bibfnamefont{F.}~\bibnamefont{Ferrer}} \bibnamefont{and}
  \bibinfo{author}{\bibfnamefont{J.~A.} \bibnamefont{Grifols}},
  \bibinfo{journal}{Phys. Rev.} \textbf{\bibinfo{volume}{D58}},
  \bibinfo{pages}{096006} (\bibinfo{year}{1998}), \eprint{hep-ph/9805477}.

\bibitem[{\citenamefont{Bedaque et~al.}(2009)\citenamefont{Bedaque, Buchoff,
  and Mishra}}]{Bedaque:2009ri}
\bibinfo{author}{\bibfnamefont{P.~F.} \bibnamefont{Bedaque}},
  \bibinfo{author}{\bibfnamefont{M.~I.} \bibnamefont{Buchoff}},
  \bibnamefont{and} \bibinfo{author}{\bibfnamefont{R.~K.}
  \bibnamefont{Mishra}}, \bibinfo{journal}{JHEP} \textbf{\bibinfo{volume}{11}},
  \bibinfo{pages}{046} (\bibinfo{year}{2009}), \eprint{0907.0235}.

\bibitem[{\citenamefont{Cerdeno et~al.}(2009)\citenamefont{Cerdeno, Munoz, and
  Seto}}]{Cerdeno:2008ep}
\bibinfo{author}{\bibfnamefont{D.~G.} \bibnamefont{Cerdeno}},
  \bibinfo{author}{\bibfnamefont{C.}~\bibnamefont{Munoz}}, \bibnamefont{and}
  \bibinfo{author}{\bibfnamefont{O.}~\bibnamefont{Seto}},
  \bibinfo{journal}{Phys. Rev.} \textbf{\bibinfo{volume}{D79}},
  \bibinfo{pages}{023510} (\bibinfo{year}{2009}), \eprint{0807.3029}.

\bibitem[{\citenamefont{Pukhov}(2004)}]{Pukhov:2004ca}
\bibinfo{author}{\bibfnamefont{A.}~\bibnamefont{Pukhov}}
  (\bibinfo{year}{2004}), \eprint{hep-ph/0412191}.

\bibitem[{\citenamefont{Cirelli et~al.}(2008)\citenamefont{Cirelli,
  Franceschini, and Strumia}}]{Cirelli:2008id}
\bibinfo{author}{\bibfnamefont{M.}~\bibnamefont{Cirelli}},
  \bibinfo{author}{\bibfnamefont{R.}~\bibnamefont{Franceschini}},
  \bibnamefont{and} \bibinfo{author}{\bibfnamefont{A.}~\bibnamefont{Strumia}},
  \bibinfo{journal}{Nucl. Phys.} \textbf{\bibinfo{volume}{B800}},
  \bibinfo{pages}{204} (\bibinfo{year}{2008}), \eprint{0802.3378}.

\bibitem[{\citenamefont{Chang et~al.}(2008)}]{:2008zzr}
\bibinfo{author}{\bibfnamefont{J.}~\bibnamefont{Chang}} \bibnamefont{et~al.},
  \bibinfo{journal}{Nature} \textbf{\bibinfo{volume}{456}},
  \bibinfo{pages}{362} (\bibinfo{year}{2008}).

\bibitem[{\citenamefont{Abdo et~al.}(2009)}]{Abdo:2009zk}
\bibinfo{author}{\bibfnamefont{A.~A.} \bibnamefont{Abdo}} \bibnamefont{et~al.}
  (\bibinfo{collaboration}{The Fermi LAT}), \bibinfo{journal}{Phys. Rev. Lett.}
  \textbf{\bibinfo{volume}{102}}, \bibinfo{pages}{181101}
  (\bibinfo{year}{2009}), \eprint{0905.0025}.

\bibitem[{\citenamefont{Israel}(2009)}]{Israel:2009}
\bibinfo{author}{\bibfnamefont{M.~H.} \bibnamefont{Israel}},
  \bibinfo{journal}{Physics} \textbf{\bibinfo{volume}{2}}, \bibinfo{pages}{53}
  (\bibinfo{year}{2009}).

\end{thebibliography}

\end{document}